\documentclass[aps,12pt]{article}
\usepackage[tbtags]{amsmath}
\usepackage{amssymb}
\usepackage{amsfonts}
\usepackage{mathbbold}
\usepackage{hyperref}
\usepackage{txfonts}
\usepackage{esint}
\usepackage{epic}
\usepackage{eepic}
\usepackage{times}
\usepackage[matrix,arrow]{xy}
\usepackage{epsfig}
\usepackage{slashed}
\usepackage[numbers,sort&compress]{natbib}
\usepackage[amsmath,thmmarks]{ntheorem}
\newtheorem{proposition}{Proposition}
\usepackage{mathrsfs}
\DeclareMathSizes{12}{13}{9.1}{6.5}
\def \ignore#1 { {} }

\def \p {\partial}

\def \Fig#1#2#3 {
\begin{figure}
\begin{center}
\scalebox{.6}{\includegraphics{#1.eps}} \label{#1}
\end{center}
\caption{#3}
\end{figure}
}

\newcommand{\figscaled}[3]{
\begin{figure}[t]
\begin{center}
\includegraphics[#3]{#1}
\end{center}
\caption{#2} \label{#1}
\end{figure}}

\topmargin by -\headsep \textheight 9.2in \oddsidemargin -25pt
\evensidemargin \oddsidemargin \marginparwidth 1.0in \textwidth
6.8in

\def \be {\begin{equation}}
\def \ee {\end{equation}}
\def \bea {\begin{eqnarray}}
\def \eea {\end{eqnarray}}
\def \beaa {\begin{eqnarray*}}
\def \eeaa {\end{eqnarray*}}

\def\bra#1{\langle{#1}|}
\def\ket#1{|{#1}\rangle}

\newcommand{\mathsym}[1]{{}}

\theoremstyle{plain}

\theoremstyle{remark}

\begin{document}

\begin{titlepage}
\title{Recursions in Calogero-Sutherland Model Based on Virasoro Singular Vectors}
\vskip 1cm
\date{}
\maketitle
\begin{center}
\renewcommand{\thefootnote}{\fnsymbol{footnote}}

\centerline{Jian-Feng Wu \footnote{wujf@itp.ac.cn},
Ying-Ying Xu \footnote{yyxu@itp.ac.cn} ,
and Ming Yu\footnote{yum@itp.ac.cn}}

\vskip .5cm
\emph{
Institute of Theoretical Physics,\\\vspace{0.1cm} Chinese Academy of Sciences,
 Beijing, 100190, China}
\end{center}

\vskip 0.3 cm
\setcounter{footnote}{0}
\renewcommand{\thefootnote}{\arabic{footnote}}

 \abstract{

The present work is much motivated by finding an explicit way in the construction of the Jack symmetric function, which is the spectrum generating function for the Calogero-Sutherland(CS) model. To accomplish this work, the hidden Virasoro structure in the CS model is much explored. In particular, we found that the Virasoro singular vectors form a skew hierarchy in the CS model.
Literally, skew is analogous to coset, but here specifically refer to the operation on the Young tableaux.  In fact, based on the construction of the Virasoro singular vectors, this hierarchical structure  can be used to give a complete construction of the CS states, i.e. the Jack symmetric functions, recursively.
The construction is given both in operator formalism as well as in integral representation.
This new integral representation for the Jack symmetric functions may shed some insights on the spectrum constructions for the other integrable systems.}

\begin{flushleft}
\end{flushleft}

\begin{flushleft}
\end{flushleft}
\end{titlepage}
\newpage
\section{Introduction and Conclusion}

The Calogero-Sutherland (CS) Model, which is an
integrable 1d many-body system, plays important roles in many different research
areas
 in physics
and mathematics. Among them are the 2D conformal field
theories(CFTs)\cite{Awata:1995ky,Sakamoto:2004rn}, the generalized matrix models\cite{Awata:1994fz,Awata:1994xd}, the fractional
quantum hall effects(FQHE) \cite{Iso:1994ui,Azuma:1993ra,Ha(1995),Haldane07b}, and there is an even more
surprising correspondence related to the  $N=2^{\ast}$ 4D
supersymmetric gauge systems\cite{AFLT,AGT,Dijkgraaf:2009pc,Belavin:2011js,Nekrasov:2009rc,Carlsson-Okounkov,Donagi:1995cf}. The spectrum of the CS model can
be generated from the Jack polynomials\cite{Baker,Kadell92,Kadell97,Stanley,Macdonald}. From the CFT point  of view, Jack symmetric functions are
naturally the building blocks for the conformal towers, the
characters of which encode the (extended) conformal symmetries.
For instance, the Jack functions related to a given Young tableaux
are believed to be in one to one correspondence with the singular
vectors of the $W$-algebra\cite{Awata:1995ky}, which reflects the hidden
$W_{1+\infty}$ symmetry of the CS model. Singular vectors in 2d CFT are the
keys to the calculation of the correlation functions in CFT and may also
reveal important physical properties of the CS model. Unfortunately,
on the  CFT side, it is not clear how to relate the
construction of the secondary states in the conformal tower to that of the Jack functions, except for some simple cases, i.e. the Virasoro singular vectors \cite{Yamada95}.
On the 2d CFT side, the calculation of conformal blocks is based on the conformal
Ward-identities,
$$[L_n, V_h(z)]=(z^{n+1}\partial_z +(n+1)h z^n )V_h(z).$$ And the calculation is carried out
perturbatively level by level \cite{Zamolodchikov82,Zamolodchikov84}. In some special cases, the
decoupling of the Virasoro null
 vectors can be implemented as differential equations for the conformal blocks. For the general
 case,  recursion relations have been proposed by Zamolodchikov
on the meromorphic structures of the conformal blocks either in complex $c$-plane or $h$-plane.
However, it remains unclear (to us) how to construct the basis vectors in the given conformal tower by  making use of the Zamolodchikov's recursion formulae explicitly.
In contrast, there are various ways in the explicit constructions of the Jack polynomials, each follows different strategies. For example, there are two integral representations. The one given in \cite{Awata:1994zr} is based on the $W_{1+\infty}$ symmetry hidden in the CS model, the other, by the authors of \cite{Okounkov97a}, starts from the so called shift Jack polynomials. There is also an operator formalism \cite{Lapointe(1995),Polychronakos:1992zk} based on the Dunkel (exchange) operators.
However, here we follow  a different strategy  in constructing the
Jack polynomials which are intrinsically related to the singular
vectors in the Virasoro algebra, and present  a new recursion
relation derived from the construction. We also do not need to
invoke the hidden $W$-algebra. In fact there is a hidden Virasoro structure in the Hilbert space of the CS model which can be used recursively in our construction. To be more specific, the ket states in the Fock space
realization of the CS model is
mapped a la Feigin-Fuks-Dotsenko-Fateev Coulomb gas
formalism \cite{Dotsenko:1984ad}\cite{Feigin:1981st} to the singular vectors  of the Virasoro algebra and its skew hierarchical descendants. The construction of the singular vectors defines a
new recursion relation for the Jack functions and finally leads to
a new integral representation which differs from the one appeared in
\cite{Awata:1994zr,Okounkov97a}. Hence our  approach may supply
new insights in dealing with CS model or more general integrable
systems.

The structure of this article is organized as following. In section 2, we
review some useful properties of the Jack polynomials and the CS model. In
section 3, we review the construction of the Virasoro singular vectors
which are related to the Jack polynomials with rectangular Young
tableaux. It should be stressed again that the Virasoro symmetry is hidden in the ket (or bra)  Hilbert space only, and is not the true symmetry in the CS model. The reason is that the prescribed Hermiticity in the CS model is not respected by the conjugation of the hidden Virasoro algebra. I.E., the Virasoro structure in the bra and ket Hilbert spaces, respectively, are not related by the Hermitian conjugation in the CS model. Our main results are given in sections 4 and 5. In section 4, we propose a skew (recursion) formula for generating new Jack functions stating from the simple ones which inevitably involve the Virasoro singular vectors,
or equivalently, the Jack functions of the rectangular graph. Our proof of the skew (recursion) formula can be considered as an operator formalism generalization of its counterpart for the Jack symmetric polynomials found by Kadell in \cite{Kadell97}.
The basic skew relation is further developed recursively in section 5. This can be made explicit first in the operator formalism, based on which we develop
a new integral representation for the Jack symmetric functions associated with any generic Young tableaux. One immediately sees the  advantages of our operator formalism of the skew (recursion) formula over the one proposed by Kadell.
Since our formalism does not depend on explicitly the number of argument variables $\{z_i\}$ for $i=1, \cdots , N$, so the recursion is done without worrying the change of the total number of arguments. Finally, the integral representation of the Jack symmetric polynomials which depend explicitly on  finite number of variables $\{z_i\}$,  for $i=1, \cdots , N$, are  presented as a by-product.

\section{Jack Polynomials and Calogero-Sutherland Model}

Now we review the Jack polynomials and the Calogero-Sutherland model.
The Jack polynomials can be viewed as a special one parameter
generalization of the Schur polynomials\cite{Macdonald} and the Jack symmetric function is the large N limit of the Jack polynomials. For physicists, the most familiar integrable
system which involves Jack symmetric functions as its spectrum
functions is the Calogero-Sutherland model. This model is an integrable
system and shows a great deal of interesting aspects, such as
duality, conformal invariance, and even the combinatorial property
of the spectrum etc. An elementary introduction can be found, for
instance, in \cite{Ha(1995)} and further studies
in \cite{Stanley} and \cite{Macdonald}. Here we only review some basics of the Jack polynomials.
\subsection{Partitions and Jack Polynomials}
Given a partition:
\(\lambda=(\lambda_1,\lambda_2,\dots,\lambda_n)\),
\(\lambda_1\geq\lambda\geq\cdots\geq\lambda_n\) ,$n\equiv l(\lambda)$, one defines the
related Jack  polynomial as the basis function for the symmetric
homogeneous polynomials in $N$ variables $\{z_i\}, i=1,2,...,N$, of
degree $|\lambda|=\sum \lambda_i$
\begin{equation}
J_{\lambda}(\{ z_i \})= \sum_{\lambda^\prime\leq\lambda ,\, l(\lambda^\prime)\leq N} C_{\lambda}^{ \lambda^\prime }z^{ \lambda^\prime  }\,,
\end{equation}
which should satisfy the second order differential equation:
\bea\label{JackDE}
H_J J_{\lambda}& = &E_\lambda J_{\lambda},\,\,\,\,\, H_J = H_J^0+\beta H_J^I\\
H_J^0& = &\sum_i^N\left(z_i\partial_{z_i}\right)^2,\,\,\,\,H_J^I
=\sum_{i<j}\dfrac{z_i +
z_j}{z_i-z_j}(z_i\partial_{z_i}-z_j\partial_{z_j}) \,, \eea here
\begin{equation}
z^{\lambda}:= \sum_{P} z_{P(1)}^{\lambda_1}\cdots z_{P(N)}^{\lambda_N}\,,
\end{equation}
$$\lambda^\prime\leq\lambda \Rightarrow \sum_{i=1}^{j} \lambda^\prime_i\leq  \sum_{i=1}^{j} \lambda_i, \, for \, j=1,2,\cdots l(\lambda')$$
 $P$ means the permutations of $N$ objects. We have also defined $\lambda_i=0$ for $i>l(\lambda)$ and \(\lambda_1+\lambda_2+\cdots+\lambda_{l(\lambda)}=|\lambda|\), $|\lambda|$ is the level of the partition. $\lambda$ can be represented graphically as a Young tableau $\lambda = \{(i,j)|1\leq i\leq l(\lambda)\},1\leq j\leq \lambda_i$. And the corresponding transposed Young tableau is represented as $$\lambda^t=(\lambda^t_1,\lambda^t_2,\dots,\lambda^t_{\lambda_1})\Rightarrow \lambda=(\lambda_1,\lambda_2,\dots,\lambda_{\lambda^t_1}).$$  It is clear that $l(\lambda)\equiv \lambda^t_1$.
We shall see later that the defining differential equation can be derived
from the CS Hamiltonian.

The Jack polynomials can be generated by the power sum symmetric polynomials as well:
\(p_l= \sum _{i=1}^N z_i^l\)
\begin{gather}
J_{\lambda}(p)=\sum_{|\lambda^\prime|=|\lambda|} d_{\lambda}^{\lambda^\prime} p_{\lambda^\prime},\ \ \ p_{\lambda^\prime} =p_{\lambda^\prime_1}p_{\lambda^\prime_2}\cdots p_{\lambda^\prime_m},\ \ \ d_{\lambda}^{1^{|\lambda|}}=1  \nonumber
\end{gather}
when $N$ large, $J_\lambda$ spans the Hilbert space of free
oscillators, and each power sum  symmetric polynomial behaves as a single oscillator. By using the
conjugacy class representation of the Young tableau $\lambda =
\{i^{m_i}\},~~i=1,2,\dots$, where $m_i$ means the multiplicity  of the rows
of $i$ squares in Young tableau $\lambda$, the normalization
of the power sum symmetric polynomial is derived from that of the oscillators,

\begin{eqnarray}\langle  p_{\lambda},p_{\lambda'}\rangle &=&\langle  \frac {a_{\lambda}
a_{-\lambda'}}{\beta^{\frac{1}{2} l(\lambda)} \beta ^
{\frac{1}{2} l(\lambda')}}\rangle
\\\nonumber &=&\delta_{\lambda \lambda'}  i^{m_i}m_i !
\beta^{-l(\lambda)}\\\nonumber \langle a_n
a_{-m}\rangle&=&\delta_{n,m} n,\,\,\, \beta=k^2\,,
\end{eqnarray}

In fact, the above normalization is consistent with that of Jack symmetric functions\cite{Ha(1995),Stanley}.
 The normalization of the Jack polynomials is derived from that of
the wave function in the CS model: \bea\label{NormHa} \left(\prod_i^N \int_0^\pi dx_i\right)
J_{\lambda^\prime}(p^{\ast})J_{\lambda}(p)\prod_{i<j} |z_i-z_j|^{2\beta} = \Gamma_N^2 \delta_{ \lambda,\lambda^\prime} j_\lambda
\dfrac{\bar{A}_{\lambda,N}}{\bar{B}_{\lambda,N}}\,,\eea here
\bea j_{\lambda}&=& A_{\lambda}^{1/{\beta}}B_{\lambda}^{1/\beta},\, \, \, z_i=e^{2ix_i}\\\nonumber
A_{\lambda}^{1/\beta}&=& \prod_{s\in
\lambda}\left(a_{\lambda}(s)\beta^{-1} +
l_{\lambda}(s)+1\right),\,\,\,
B_{\lambda}^{1/\beta} =
\prod_{s\in\lambda}\left((a_{\lambda}(s)+1)\beta^{-1}+l_{\lambda}(s)\right)\,\eea
$a_{\lambda}(s)$ and $l_{\lambda}(s)$ are called arm-length and
leg-length of the box $s$ in the Young tableau $\lambda$:
\[a_{\lambda}(s)=\lambda_i-j, \,\,\,\,\,l_{\lambda}(s)=\lambda^t_j-i,\] $\lambda^t_j$ is
the $j$-th part of the partition related to the transposed Young
tableau $\lambda$.
$$
\bar{A}_{\lambda,N} = \prod_{s\in \lambda}\left( N+a'_{\lambda}(s)/\beta
-l'_{\lambda}(s)\right),\,\,\,\, \bar{B}_{\lambda,N} =\prod_{s\in \lambda}\left(
N+(a'_{\lambda}(s)+1)/\beta-(l'_{\lambda}(s)+1) \right),$$ $a'_{\lambda}(i,j) = j-1$,
$l'_{\lambda}(i,j)= i-1$ denote the co-arm-length and co-leg-length for the
box $s=(i,j)$ in Young tableau $\lambda$. $\Gamma_N^2 \equiv
\pi^{N}\dfrac{\Gamma(1+N\beta)}{\Gamma^N(1+\beta)}$
is the normalization of the ground state. When $N\to\infty$, and
after dividing out the ground state normalization $\Gamma_N^2$, we
obtain the normalization for the Jack functions:
\bea\label{orthpoly}\langle J_\lambda, J_{\lambda'}\rangle &=& \int
J_{\lambda'} (p^\ast)J_\lambda (p)\dfrac{\prod_{i<j}|z_i-z_j|^{2\beta}  }{\Gamma_N^2}\prod_{i=1}^N
dx_i\mid_{N\to\infty} = \delta_{\lambda,\lambda'} j_\lambda\eea

We shall see in the next section that eq.(\ref{NormHa}) implies the following integral formula
\be\label{intformJack}J_\lambda(\dfrac{a_-}{k}) = \int e^{k\sum_{n>0}
\frac{a_{-n}}{n}p_n} J_\lambda(p^\ast)\dfrac{\prod_{i<j} |z_i-z_j|
^{2\beta} }{\Gamma_N^2} \prod_{i=1}^N dx_i\,.\ee

Now we shall clarify some notations used in this work. $J_\lambda(p)$ means Jack polynomials in the power sum polynomial basis,  $J_\lambda(\frac{a}{k})$
the annihilation operator valued Jack symmetric function, i.e. with the substitution $p_n \rightarrow \dfrac{a_n}{k}$, and $J_{-\lambda}\equiv J_\lambda(\frac{a_-}{k})$ the creation operator valued Jack symmetric function, i.e. with the substitution $p^\ast_n \rightarrow \dfrac{a_{-n}}{k}$.
\subsection{Calogero-Sutherland Model}
The CS model is introduced for studying Coulomb interacting
electrons distributed on a circle. The Hamiltonian for this system
can be written as\footnote{For convenience, we set the circumference
of the circle as $L=\pi$}:
\begin{eqnarray}
H_{CS}&=&-\sum_{i=1}^N\frac{1}{2}\partial_i^2+\sum_{i<j}\frac{\beta(\beta-1)}{
\sin^{2} (x_{ij})}
\end{eqnarray}
here  \(\partial_i=\partial_{x_i}, \,\hbar^2/8m =1\), $\beta=k^2$,
$k$ is the charge of the N identical electrons. This is an exact
solvable system. However, let's consider another auxiliary
Hamiltonian which is positive definite and differs from the original
one only by a shift of the constant ground state energy
\begin{eqnarray}\label{Hamiltonian}
H&=&-\frac{1}{2}\sum_{i=1}^N(\partial_i+\partial_i
\ln\prod_{l<j}\sin^{\beta}x_{lj})(\partial_i-\partial_i
\ln\prod_{l<j}\sin^{\beta}x_{lj})\\\nonumber&=&-\frac{1}{2}\sum_{i=1}^N\partial_i^2+\beta(\beta-1)\sum_{i<j}\frac{1}{\sin^2
x_{ij}}-\frac{1}{6}\beta^2N(N+1)(N-1)\\\nonumber&=&H_{CS}-E_{0}\,,\\\nonumber
E_{0}&=&\frac{1}{6}\beta^2N(N+1)(N-1)\,.
\end{eqnarray}
In going from the first line to the second line of eq.
(\ref{Hamiltonian}), we have used the identity
\begin{eqnarray}
&&\sum_{i,j\neq k}\cot x_{ij}\cot x_{ik}+\sum_{j,i\neq k}\cot
x_{ji}\cot x_{jk}+\sum_{k,i\neq j}\cot x_{ki}\cot x_{kj}\\\nonumber
&=& \sum_{i,j\neq k} \frac{-\cos x_{ij}-\cos x_{ik}\cos x_{jk}}{\sin
x_{ik}\sin x_{jk}}\\\nonumber&=&\sum_{i,j\neq k}(-1)=-N(N-1)(N-2)\,.
\end{eqnarray}
where $x_{ij}\equiv
x_i-x_j$. It is also convenient to define $z_i=e^{2i x_i}$ for later use.
 The ground state should be a solution to
the eigen-equation:
\[H_{CS}\psi_0 =E_0\psi_0,\] and can be easily read out:
\[\psi_0=\prod_{i<j}(2\sin^{\beta}x_{ij})\, ,\]
where the factor of 2 is included for normalization reason.
If one defines the excited state as
\[\psi_\lambda=J_{\lambda}\psi_0\,,\] then it can be shown that this state actually satisfies the energy eigen-equation:
\begin{eqnarray}
H\psi_{\lambda}&=&H\psi_0
J_{\lambda}(p)=\psi_0(\psi_0^{-1}H\psi_0)J_{\lambda}(p)
\\\nonumber&=&2\psi_0 H_J J_{\lambda}(p)=2\psi_0 E_{\lambda}
J_{\lambda}(p)\,,\\\label{HamiltonJ}
H_J&=&\frac{1}{2}\psi_0^{-1}H\psi_0 =
-\frac{1}{4}\sum_{i=1}^N (\partial_i+2\beta\sum_{j\neq i}\cot
x_{ij})\partial_i\,,\end{eqnarray} thus the eigen-equation can be
rewritten as
\begin{eqnarray}\label{HJ}H_JJ_{\lambda}
&=&\left[-\dfrac{1}{4}\sum_{i=1}^N\partial_i^2-\dfrac{1}{2}\beta \sum_{i< j} \cot x_{ij}
(\partial_i -
\partial_j)\right] J_{\lambda} = E_{\lambda}J_{\lambda}\,.\end{eqnarray}  We see that
this  coincides with the defining differential equation
eq.(\ref{JackDE}) for the Jack polynomials. The eigenstates of $H_J$ in the form of eq.(\ref{HamiltonJ}) and(\ref{HJ}), means that $H_J$ is triangular with respect to the symmetric monomials.
\[J_\lambda \sim \left(\prod_{i=1}^{l(\lambda)}z_i^{\lambda_i}+symmetrization\right)+daughter\,\,\, terms\,.\] Here the
daughter
terms are the symmetrized monomials associated with Young tableau $\lambda^\prime <\lambda$. That is to say, given a Young tableau $\lambda$, one can squeeze the
partition $\{\lambda\}$ to other partitions by moving squares in $\lambda$ downwards to get new Young
tableaux. These terms actually reflect the triangular property of
the interaction $H_J^I$ as in eq.(\ref{JackDE}).
fig.{\ref{squeezing}} gives an example of squeezing.
\figscaled{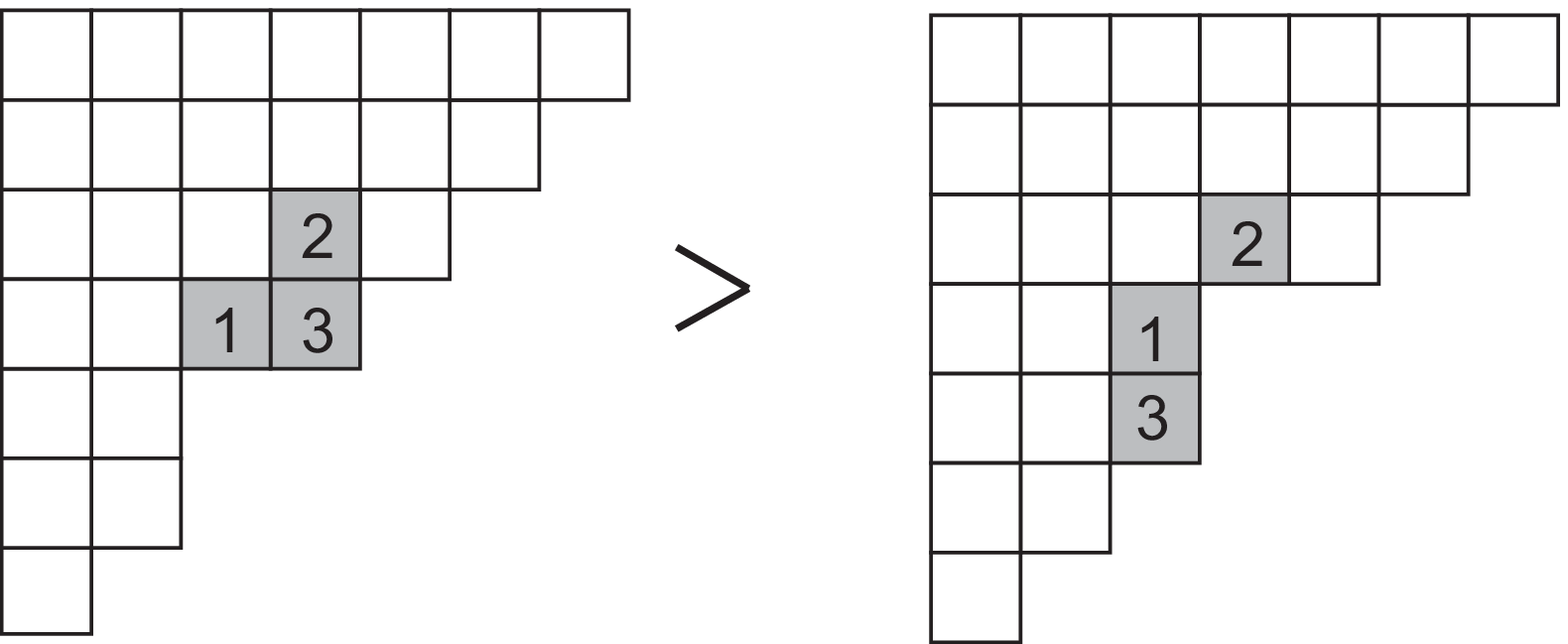}{An example for squeezing  Young tableau,
where the square 3 has been squeezed downward to form a different
Young tableau.}{height=2in}  It is easy to
read out the eigenvalue of $H_J$ which can be read off from the
diagonal value of the leading term of the eigenstate
\begin{eqnarray}\label{Elambda}\nonumber
E_{\lambda}&=&-\frac{1}{4}\sum_i^N(-4\lambda_i^2)-constant\,\,
term\,\, part\,\, \left(\frac{1}{2}\sum_{i< j}i\beta\frac{(z_i+z_j)}{(z_i-z_j)}
\frac{(z_i^{\lambda_i}z_j^{\lambda_j}-z_j^{\lambda_i}z_i^{\lambda_j})}
{(z_i^{\lambda_i}z_j^{\lambda_j}+z_j^{\lambda_i}z_i^{\lambda_j})}(2i\lambda_i-2i\lambda_j)\right)\\\nonumber&=&
\sum_i^N\lambda_i^2+\beta\sum_{i<j}(\lambda_i-\lambda_j)\\&=&
\sum_i^N\lambda_i^2 +
\beta\sum_i^N(N-2i+1)\lambda_i\,.
\end{eqnarray}Here in the last step of eq.(\ref{Elambda}), we have used the fact that \beaa\sum_{i<j}^N (\lambda_i- \lambda_j)& =& (N-1) \lambda_1
+(N-2-1)\lambda_2+\cdots
+(N-2i+1)\lambda_i\cdots\\&=&\sum_i^N(N-2i+1)\lambda_i\,.\eeaa
Here, since we are concerning ourselves to all the Young tableaux $\lambda^\prime \leq\lambda$ with the restriction $l(\lambda^\prime) \leq N$,
so we redefine $\lambda$ as well as $\lambda^\prime$  to include the trailing null parts such that $l(\lambda^\prime) =l(\lambda) =N$.
Actually, eq.(\ref{Elambda}) can be written as the following more
compact formula: \bea E_{\lambda} &=& k\left(k^{-1}||\lambda||^2-
k||\lambda^t||^2 +kN |\lambda|\right)\\\nonumber ||\lambda||^2 &\equiv&
\sum_i^N\lambda_i^2,\,\,\,\,||\lambda^t||^2 \equiv
\sum_i^N(\lambda_i^t)^2,\,\,\,\,|\lambda| \equiv
\sum_i^N\lambda_i. \eea
\subsection{Second Quantized Form}
In fact, the second quantized form of the CS model can be realized as a theory of 2D scalar field $\varphi(z)$\cite{Awata:1994xd}.
In the corresponding
 CFT,  \(\varphi(z)\) is a scalar defined on the unit circle but can be analytically continued to complex plane. The vertex operator
 for CS model reads
\begin{eqnarray}
V_k(z)&=&:e^{k \varphi(z)}:\\\nonumber \varphi(z)&=&q+plnz +\sum_{n\in
z,n\neq0}\frac {a_{-m}}{m} z^m
\\\nonumber \langle\varphi(z)\varphi(w)\rangle &=&\log (z-w)\,,
\end{eqnarray}
here $$[a_n,a_m]=n\delta_{n+m,0} ,\,\,\,\,[p,q]=1, \,\,\,\,\,
\varphi(z)^\dagger =- \varphi(z)\,.$$ It is easy to show that the ground
state of the CS model can be written as the holomorphic part of the
correlation function in conformal field theory \bea\label{vertop} \langle
k_f|V_k(z_1)\cdots V_k(z_n)|k_i\rangle &=&\prod_{i<j}^N
(z_i-z_j)^{k^2}\prod_{j=1}^N z_j^{k_i \cdot k}. \eea If one choose $a_0|k_i\rangle=k_i|k_i\rangle$,\, \, $k_i=\frac{k}{2}(1-N)$, the correlation function reproduces the
ground state of CS model up to a phase  factor \footnote{For
simplicity, we drop this phase factor in the following context.}:
\begin{equation} \prod_{i<j}^N (z_i-z_j)^{k^2}\prod_{j=1}^N z_j^{k_i
\cdot k} = (i)^{k^2\frac{N(N-1)}{2}} \prod_{i<j}
(2\sin^{\beta}x_{ij}).
\end{equation}
Noticing that the ground state actually comes from the contraction
of the vertex operators, then we can define the state $|\psi\rangle$
as
\begin{eqnarray}\label{state}
|\psi\rangle&=&\prod_{j=1}^{n}V_k(z_j)|k_i\rangle\sim\prod_{i<j}(2\sin^{\beta}x_{ij}):\prod_{j=1}^N V_k(z_j):|k_i\rangle\\\nonumber
&=&\psi_0(x_i)e^{k\sum_{n>0}a_{-n}p_n/n}|\frac{N+1}{2}k\rangle\equiv\psi_0(x_i)V_k^{(-)}(p)|\frac{N+1}{2}k\rangle\ \, ,
\end{eqnarray} with the action of the CS Hamiltonian, one obtains:
\begin{eqnarray}
\nonumber \frac{1}{2}H|\psi\rangle&\sim&-\frac{1}{2}\psi_0\sum_i(\partial_i +2\beta \sum_{j\neq
i}\cot x_{ij})\partial_i V_k^{(-)}(p)|k_i\rangle\\\nonumber&=&\psi_0
\left(\sum_{n>0}a_{-n}a_n(\beta N +n(1-\beta))+k\sum_{n,m>0}
(a_{-n-m}a_n a_m +a_{-n}a_{-m}a_{n+m})\right)V_k^{(-)}(p)|k_i\rangle.
\end{eqnarray}

Then we get the second quantized form of the Hamiltonian,
\begin{eqnarray}\label{OperatorH}
\hat{H}:= \sum_{n>0}a_{-n}a_n(\beta N +n(1-\beta))+k\sum_{n,m>0}
(a_{-n-m}a_n a_m +a_{-n}a_{-m}a_{n+m}),
\end{eqnarray}
and the wave function in coordinate space:
\begin{equation}
\psi_{\lambda}(\{z_i\})=\langle k_f|
J_{\lambda}(a/k)\psi_0 V_k^{(-)}(p)|k_i\rangle
\end{equation}
does satisfy the eigen-equation
\begin{eqnarray}
H\psi_{\lambda}&=&\langle k_f|
J_{\lambda}(a/k)H\psi_0V_k^{(-)}(p)|k_i\rangle\\\nonumber &=&2\langle k_f|
J_{\lambda}(a/k)\psi_0 \hat{H}V_k^{(-)}(p)|k_i\rangle\\\nonumber&=&2E_{\lambda}\psi_{\lambda}\,,
\end{eqnarray}
provided the following defining operator equation for the Jack functions is satisfied.
\be
\langle 0|J_\lambda(a/k)\hat{H}=\langle 0 |J_\lambda(a/k)E_\lambda\,.
\ee

\subsection{Duality Relation}
 Since for
CS system $a_0 = \dfrac{1}{2}(N+1)k$, the Hamiltonian,
eq.(\ref{OperatorH}) can be written in a more compact form
as\footnote{For convenience, we neglect the summation symbols, one
can recover them whenever one needs.} \bea\hat{H}  &=& \sum_{n,m>0}k(a_{-n}a_{-m}a_{n+m} +a_{-n-m}a_n a_m) + \sum_{n>0}(2a_0 a_{-n}a_n+(1-\beta)na_{-n}a_{n}- \beta a_{-n} a_n)
\\\nonumber &=&\frac{1}{3} k\oint(z\partial_{z}\varphi(z)-a_0)^3\dfrac{d z}{2\pi iz}+\sum_{n>0}2a_0 a_{-n}a_n+\sum_{n>0}(1-\beta)n a_{-n} a_n - \beta a_{-n}a_{n}\\\nonumber
 &\equiv& k \left(\hat{H'}(k) +(2 a_0-k)a_{-n} a_n\right)\,.
\eea

There exists an explicit duality relation which can be read off as
follows. First, we have the non-zero mode part of $\hat{H}$ as
\bea\label{dualhamiltonian}\hat{H'}(k) = \sum_{n,m>0}(a_{-n} a_{-m} a_{n+m} +
a_{-n-m} a_n a_m )+\sum_{n>0}(k^{-1}-k)n a_{-n}a_n\,,\eea it has an apparent
symmetry \bea k^{-1}\leftrightarrow -k\,, \eea namely, let
$\tilde{k}= -k^{-1}$ \bea \hat{H'}(\tilde{k}) = \hat{H'}(k)\,.\eea

Now we shall show that $k\rightarrow \tilde{k}$ sends Young tableau
$\lambda$ to its dual diagram (transposed
diagram) $\lambda^t =\{\lambda_1^t,\lambda_2^t,\dots,\lambda_N^t\}$. Since $\hat{H'}(k)$ acts
on Jack function gives \bea
\hat{H'}(k) |J_\lambda\rangle  &=& E^{(k)}_Y|J_{\lambda}\rangle  \\
E^{(k)}_\lambda &=& \sum_i \left(\lambda_i^2k^{-1} - (2i -1)\lambda_i
k\right)\\\nonumber&=&\sum_i\left({\lambda^t}_i^2\tilde{k}^{-1}-(2i-1)\lambda^t_i
\tilde{k}\right) = E^{(\tilde{k})}_{\lambda^t} \eea Here we have used the following identity \[\sum_{i=1}^{l(\lambda)}(2i-1)\lambda_i = \sum_{j=1}^{l(\lambda^t)}(\lambda^t_j)^2.\] We now conclude that
\be\label{kdual}
J_\lambda^{1/\beta}(\dfrac{a_-}{k})=(\dfrac{-1}{\beta})^{|\lambda|}
J^{\beta}_{\lambda^t}(-k a_-)\,.\ee Finally, notice that the
inclusion of of the $a_0$ part of $\hat{H}$,  $(2 a_0-k)a_n
a_n=Nk|\lambda|$, will not change eq.(\ref{kdual}), only the eigenvalue of $\hat{H}$ is different
on the two sides of eq.(\ref{kdual}).

\subsection{Generating Function and Skew-Fusion Coefficient}

The bra state $|\psi\rangle$ defined in eq.(\ref{state}), is actually the second
quantized form of the wave packet, which is a linear superposition
of the incoming energy eigenfunctions defined on the unit circle.
The superposition coefficients are understood as the creation
operators creating incoming energy eigenstates. To see this, from
eq.(\ref{kdual}) and orthogonality condition, eq.({\ref{orthpoly}}), we can expand , \bea\label{Frobenius}
\exp(k\sum_{n>0}\dfrac{a_{-n}}{n} p_n) &=& \sum_\lambda \dfrac{J_{\lambda}^{1/\beta}(
\frac{a_-}{k})}{j_\lambda^{1/\beta}}J_{\lambda}^{1/\beta} (p)
\Rightarrow \\\nonumber \exp(-\dfrac{1}{k}\sum_{n>0}\dfrac{a_{-n}}{n} p_n)
&=&\sum_{
\lambda}\dfrac{J_{\lambda}^{1/\beta}(\dfrac{a_-}{k})(-)^{|\lambda|}P_{\lambda^t}^{\beta}(p)
}{A_{\lambda}^{1/\beta}}\,.\eea Here, the last equality in eq.(\ref{Frobenius}) comes from the
duality relation, eq.(\ref{kdual}), and we have also  defined
$$P_\lambda^{\beta}(p) = \frac{J_{\lambda}^{\beta} (p) }{\prod_{s\in \lambda}(a_\lambda(s)\beta+l_\lambda(s)+1)}= \frac{J_\lambda^\beta(p) }{A_\lambda^\beta}\, , $$
which is proportional to the Jack polynomials but normalized differently,
\begin{equation}
P_\lambda^{1/\beta}(p)=z^{\lambda} +\sum_{\lambda^\prime < \lambda} m_{\lambda}^{\lambda^\prime}z^{\lambda^\prime}\,,
\end{equation}
Similarly $\psi^\dagger$ creates outgoing states, \bea
\exp(\dfrac{1}{k}\sum_{n>0}\frac{a_n}{-n} p_{-n})
&=&\sum_\lambda \dfrac{J_\lambda^{1/\beta}(\dfrac{a}{k})(-)^{|\lambda|}
{P_{\lambda^t}^{\beta}} (p^\ast) }{A_{\lambda}^{1/\beta}} \,;\\
\exp(-k\sum_{n>0}\dfrac{a_n}{-n} p_{-n}) &=&
\sum_\lambda\dfrac{J_\lambda^{1/\beta}(\dfrac{a}{k})J^{1/\beta}_\lambda (p^\ast)
}{j_\lambda^{1/\beta}}\,. \eea
Besides being a wave packet, the state $|\psi\rangle$ is also a
coherent state which is the eigenstate for all the annihilation
operators ${J_\lambda}(k^{-1}a)$ with eigenvalue $J_\lambda(p)$ for
each Young tableau $\lambda$. To show that the r.h.s of eq.(\ref{Frobenius}) is actually a coherent state, we need define a 3-point function\cite{Stanley}
$\langle  J_{\mu} J_{\nu}J_{-\lambda}\rangle\equiv g_{\mu\nu}^\lambda\Rightarrow
J_{\mu}(p)J_{\nu}(p) = \sum_{\lambda}g_{\mu\nu}^\lambda
j_\lambda^{-1}J_\lambda(p)\,, $ and $J_\mu J_{-\nu}|\rangle=\sum_{\lambda}g_{\mu\lambda}^\nu j_{\lambda}^{-1} J_{-\lambda}|\rangle:=J_{-\nu/\mu}|\rangle\,$
is called the skew Jack symmetric function.
Hence we have the following equation. \bea
&&J_\mu|\psi\rangle =\sum_\nu J_\mu\dfrac{J_{-\nu}J_\nu(p)}{j_\nu}|k_i\rangle=
\sum_{\lambda,
\nu}g_{\lambda,\mu}^{\nu}\dfrac{J_{-\lambda}}{j_{\lambda}}\dfrac{J_{\nu}(p)}{j_{\nu}}|k_i\rangle
=\sum_{\lambda}
\dfrac{J_{-\lambda}J_\lambda(p)}{j_{\lambda}}J_{\mu}(p)|k_i\rangle\\\nonumber
&&=J_\mu(p)|\psi\rangle \eea
In general, the fusion coefficient $g_{\mu\nu}^{\lambda}$ is not a simple expression. However, if a rectangular Young tableau is involved, then $g_{ \lambda,s^r/\lambda}^{s^r}$  can be derived from the
generating function, eq.(\ref{Frobenius}) and the normalization
condition, eq.(\ref{NormHa}). One gets\footnote{In this particular case, $s^r/\lambda$ is taken to represent the Young tableau as in fig.\ref{skewyoung} and $J_{s^r/\lambda}$ the corresponding Jack function associated with it.}
 \bea\label{fusioncoeff}
J_{\lambda} J_{-s^r}|0\rangle &=& A_\lambda \oiint
V_k^{(-)}(p)P_\lambda(p) z_i^{-s-1}dz_i
\dfrac{\prod_{i<j}|z_i-z_j|^{2\beta}}{\Gamma_r^{2}}B_{s^r}\ket{0}\\\nonumber&=&
A_\lambda J^{1/\beta}_{s^r/\lambda}(\dfrac{a_-}{k})
\dfrac{\bar{A}_{s^r/\lambda},r}{A_{s^r/\lambda}\bar{B}_{s^r/\lambda,r}}B_{s^r}\ket{0}\, ,
\eea
In deriving this\footnote{We drop the superscript $1/{\beta}$ of $A_{\lambda}^{1/\beta}$ etc and add it explicitly when necessary.}, use has been made of the following identity \cite{Kadell97},
\be\label{Kadell} P_\lambda(p) z_i^{-s} = P_{s^r/\lambda}(p^\ast)\, . \ee
Thus \bea \label{coeff1}g_{\lambda,s^r/\lambda}^{s^r}j^{-1}_{s^r/\lambda} &=& A_\lambda
\dfrac{\bar{A}_{s^r/\lambda,r}}{A_{s^r/\lambda}\bar{B}_{s^r/\lambda,r}}B_{s^r,r}\\\nonumber&=&
\dfrac{B_{s^r,r}\bar{A}_{\lambda,r}}{\bar{B}_{s^r/\lambda,r}}.\eea In reaching the last line in the above equation,
we have used the following interesting identity:
\bea\label{identity1} A_{\lambda}\bar{A}_{s^r/\lambda,r} = A_{s^r/\lambda} \bar{A}_{\lambda,r}.
\eea This identity can be proven diagrammatically by moving squares in the Young tableaux. The detailed presentation on this diagrammatic proof will appear elsewhere \cite{WuXuYu2011c}.

Another example which involves the skew Jack function is of two sets of oscillators.
Let's  consider the following expansion:
\be\label{expansion1}
\exp(\sum_{n>0}k\dfrac{(a_{-n}+\tilde{a}_{-n})p_n}{n})=\sum_{l(\lambda)\leq N}\dfrac{J_{\lambda}(\dfrac{a_{-}+\tilde{a}_{-}}{k})J_\lambda(p)}{j_\lambda} \,.\ee
One can also expand $\exp(\sum_{n>0}k\frac{(a_{-n}+\tilde{a}_{-n})p_n}{n})$
in another way,
\bea\label{expansion2}
&&\exp(\sum_{n>0}k \dfrac{a_{-n}p_n}{n})\exp(\sum_{n>0}k\dfrac{\tilde{a}_{-n}p_n}{n})\\\nonumber
&=&
\sum_{\mu,
\nu} \dfrac{J_{\mu}(\dfrac{a_{-}}{k})}{j_{\mu}}J_{\mu}(p)\dfrac{J_{\nu}(\dfrac{\tilde{a}_{-}}{k})}{j_{\nu}}
J_{\nu }(p)\\\nonumber&=&\sum_{|\mu|+|\nu|=|\lambda|}
\dfrac{J_{\mu }(\dfrac{a_{-}}{k})}{j_{\mu }}\dfrac{J_{\nu }(\dfrac{\tilde{a}_{-}}{k})}{j_{\nu }}
\dfrac{g_{\mu  \nu }^\lambda J_\lambda(p)}{j_\lambda}\\\nonumber
&=& \sum_{\mu, \lambda}\dfrac{J_{\mu }(\dfrac{a_{-}}{k})}{j_{\mu }}J_{\lambda/\mu }(\dfrac{\tilde{a}_{-}}{k}) \dfrac{J_\lambda(p)}{j_\lambda} \,.\eea
Comparing eq.(\ref{expansion1}) and eq.(\ref{expansion2}), we find
\bea
J_{\lambda}(\dfrac{a_{-}+\tilde{a}_{-}}{k}) =
\sum_{\mu }\dfrac{J_{\mu }(\dfrac{a_{-}}{k})}{j_{\mu }}J_{\lambda/\mu }(\dfrac{\tilde{a}_{-}}{k}) \,.
\eea
Such that the skew Jack function can be obtained from the inner product
\be\label{skew}
J_{\lambda/\mu }(\dfrac{\tilde{a}_{-}}{k})=
\bra{0} J_{\mu }(\dfrac{a}{k})J_{\lambda}(\dfrac{a_{-}+\tilde{a}_{-}}{k})\ket{0}\, .
\ee
Here, $\bra{0}$ and $\ket{0}$ are the bra and ket vacuum states for the $a_n$'s only.
Eq.(\ref{skew}) turns out to be very useful when we develop a skew-recursive integral for the construction of Jack states in section 5.

\section{Virasoro Singular Vectors in Calogero-Sutherland Model}
From the discussions in the previous sections, we can see that there exist  apparent similarities between the CS
model and the Coulomb gas picture. The Coulomb gas picture endowed
with screening charges originated in \cite{Feigin:1981st} and
\cite{Dotsenko:1984ad}. This method plays an important role in the calculations of the
correlation functions in 2D conformal field theories. The conformal blocks are calculated with the insertions of the
primary vertex operators which usually ends up with a charge
deficit. In Coulomb gas
picture, such kind of charge deficit can be compensated by
sandwiching a number of conformally invariant screening charges to
make charge balanced while keeping conformal invariance intact. To
see the similarities between the CS model and the Coulomb gas picture,
notice the following:

1) For one scalar theory, we have two kinds of vertex operators which may be interpreted as
screening vertex operators with the charges
$\alpha_\pm$ in CFT and
$\pm k^{\mp}$ in CS model.

2) In both cases there are zero norm states.

3) In CFT the descendant states are generated by the Virasoro
algebra, while in the CS model the Jack symmetric functions. Both
expands a complete set of basis.

However, despite all of those similarities, we have to address some apparent dis-similarities:

1)\ \ $\alpha_+\alpha_-=-2$ while $k^{\pm}(-k)^{\mp}=-1$

2)\ \ In CFT, zero norm state exists for generic $\alpha_\pm$, while in
CS model only for $k^2\leq 0$, see eq.(\ref{NormHa})

3)\ \ In CFT the conjugate state is defined by $L_{-n}^{\dagger}=L_n$,
while in CS model $a_{-n}^{\dagger}=a_n$. The two conjugations
coincide only in the case when $k^2\leq 0 \Rightarrow c\geq 25$.

Combining the above comparison 2) and 3) we see that there is an
chance to map the two systems into each other in the case of
Liouville type CFT, provided we can solve the problem 1), i.e.
mapping between $\alpha_\pm$ and $\pm k^\mp$. It turns out that it
can be solved by introducing an additional scalar field. For
example, in AGT conjecture, an additional $U(1)$ scalar is needed to
make the comparison between Nekrasov instanton counting and the
conformal blocks of the Liouville type, where the Virasoro structure is explicitly shown (\cite{AFLT,AGT}). In that case, Jack functions are the essential ingredients in building up the desired conformal blocks. We shall postpone our discussion on this point until our next paper\cite{ShouWuYu2011} which is finishing soon. However, in the present paper, we shall restrict ourselves to the case of one set of oscillators in the operator formalism and to the case of generic $k$. In this case, we shall see that the Virasoro structure is
implicit.

\subsection{Hidden Virasoro Structure}

The existence of the Virasoro structure in the Jack symmetric function has been investigated by the authors of \cite{Awata:1995ky,Awata:1994fz}. In particular, it has been found there is a direct map between Virasoro singular vectors and the Jack functions of the rectangular Young tableau. Although it was suggested in \cite{Awata:1995ky} that  such relationship should lead to an integral representations for the Jack functions, only in some simple cases, the explicit construction was found. Starting from the next section, we shall present a complete construction for the
Jack functions based on the Virasoro null vectors and their skew hierarchies.
Here, to see how it works, we shall make some preparations. Let's rewrite the Hamiltonian $\hat{H}$ as\footnote{This
redefinition is not unitary but it makes the following
computation simpler.}
\bea\label{modifiedH}\hat{H}=\sum_{n>0}\left(\alpha_+\tilde{a}_{-n} \tilde{L}_n+(N\beta+\beta-1-\alpha_+\tilde{a}_0)\tilde{a}_{-n}\tilde{a}_n\right),\eea
{}here we have redefined \(\tilde{a}_n=\sqrt{2}a_n,
\tilde{a}_{-n}=\frac{a_{-n}}{\sqrt{2}}, n>0, \tilde{a}_0=\sqrt{2}a_0, \alpha_\pm=\pm\sqrt{2}k^{\pm 1},\alpha_+ +\alpha_- =2\alpha_0\), and the Virasoro generator
\bea\label{Virredef}\tilde{L}_n=
\frac{1}{2}\sum_{m\in \mathbb{Z}}:\tilde{a}_{m}\tilde{a}_{n-m}:-(n+1)\alpha_0\tilde{a}_n.\eea Notice
that in this convention, the Hamiltonian separates into two parts, one
for the "Virasoro part" which is proportional to $\tilde{L}_n$ and the
other part is in fact the conserved charge and is always diagonal on Jack functions and its eigenvalue  proportional to the norm of the Young tableau. It is clear that  any ``Virasoro'' singular vector $ |\chi_{r,s}\rangle$ is an eigenstate of $\hat{H}$ whose eigenvalue suggests that $|\chi_{r,s}\rangle$ is proportional to the Jack state $J_{\{s^r\}}$. Of course, The singular vector in the ``Virasoro'' sector is not singular on the CS model side, since for generic $k$, Jack functions has non-zero norm. This is because the redefinition, eq.(\ref{Virredef}),is not unitary and the conjugation in the ``Virasoro'' sector
 is not hermitian. While in the CS model, the conjugation is always Hermitian for real $k$.

 \figscaled{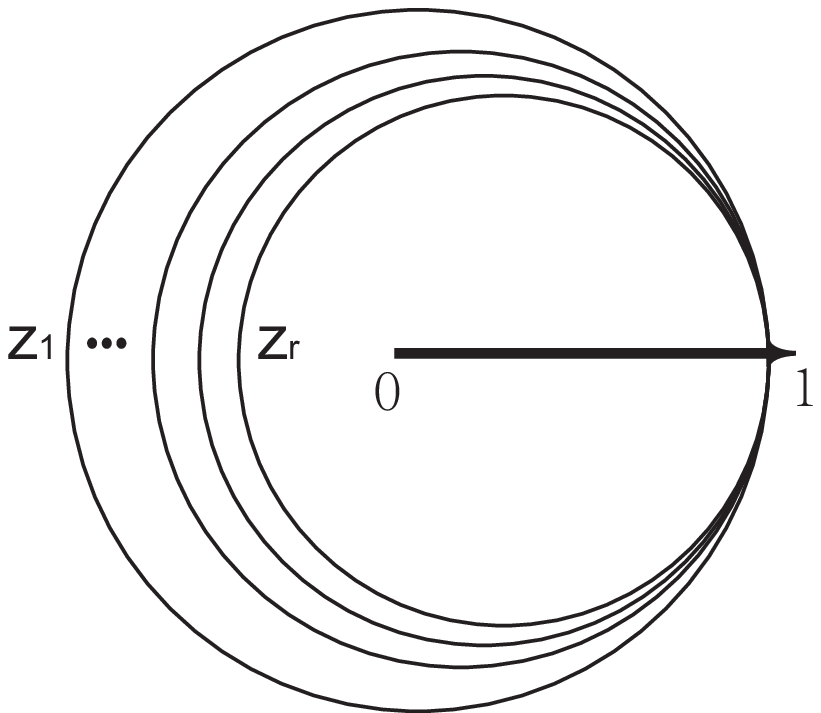}{Felder's integration contour}{height=2in}

To make the comparison more clear, we shall assume that the $a_0$ eigenvalues differ from $k_i$ defined in eq.(\ref{vertop}). Consider a general vacuum state $|p\rangle$
in the CS model, which is mapped to a highest weight state with
conformal dimension  $h_p = \frac{1}{2}p(p-2\alpha_0)$ in the Virasoro sector. The singular vectors appears when its descendant states combine themselves  into a highest weight state again. This can happen for quantized $p$
\[p=p_{r,s}\equiv\frac{1}{2}(1-r)\alpha_+ + \frac{1}{2}(1-s)\alpha_-\, .\] and at the level $rs$.
And this null vector can be constructed explicitly by making use of the fact that $h_{p_{r,s}}= h_{p_{-r,-s}}$,
 $ |\chi_{r,s}\rangle=S^r |p_{r,-s}\rangle$ which satisfies
\begin{eqnarray}
\tilde{L}_n|\chi_{r,s}\rangle &=&\delta_{n,0}( h_{p_{r,s}}+rs)|\chi_{r,s}\rangle \,
,n\geq 0\\\nonumber
\tilde{a}_0|\chi_{r,s}\rangle&=& p_{-r,-s}|\chi_{r,s}\rangle.
\end{eqnarray}
Here $S\equiv S^+=\oint V^+(z) dz\, , V^{\pm}(z)=:\exp(\alpha_\pm \tilde{\varphi}(z)):\, , \alpha_\pm =\pm\sqrt{2} k^{\pm 1}
$ are called the screening charges in the Virasoro sector. When multiple $S$' act together, we take Felder's contour \cite{Felder:1988zp} for $S^r$ (see fig.\ref{Feldercontour}). to get
\be\label{recjackint}
|\chi_{r,s}\rangle=S^r |p_{r,-s}\rangle=\oiint
\prod_{i<j}^{r}|z_i-z_j|^{2\beta}e^{k \sum_{n>0}a_{-n} p_n}
\prod_{i=1}^{r} z_i^{-s-1}dz_i|p_{-r,-s}\rangle \propto J_{-s^r}|p_{-r,-s}\rangle  \, .\ee
Notice that in the equation above we have used $a_{-n}$ instead of $\tilde{a}_{-n}$ to make the comparison with eq.(\ref{state}).

\subsection{An Example of Single Screening Charge}
The construction of the Jack states for the rectangular diagrams, as well as the null vectors of the Virasoro algebra hidden in the CS model, thus reduces to the evaluation of the multi-integrals of the Selberg type in eq.(\ref{recjackint}). Since there is no closed formula for such type of operator valued multi-integrals, we choose to discuss some simple cases here. The simplest one is the case of one screening charge for the Young tableau $\{1^n\}$. From eq.(\ref{recjackint}) and duality relation eq.(\ref{kdual}), one can verify that the state\footnote{We have dropped
the factor $\frac{1}{2\pi i}$ for convenience. We also use the label $J_{s^r}$ instead of $J_{\{s^r\}}$ for the same reason. }
\bea
|J_{1^n}\rangle &=&\oint e^{-\frac{1}{k}\sum_{m>0}\frac{a_{-m}z^m}{m}} (-1)^n n! z^{-n-1}dz|p_{-n,-1}\rangle \\\nonumber &=& \oint e^{\alpha_-\tilde{\varphi}(z)}(-1)^n n! dz|p_{-n,1}\rangle\\\nonumber &=&\oint e^{p_{-n,1}\tilde{q}}e^{z \tilde{L}_{-1}}(-1)^n n!z^{-n-1} dz|p_{1,-1}\rangle\\\nonumber
&=&e^{p_{-n,1}\tilde{q}}(-\tilde{L}_{-1})^n|p_{1,-1}\rangle
\eea
reproduces the Jack polynomial $J_{\{1^n\}}$. To take its conjugate state we have to be careful to take its Hermitian conjugation. Now let's workout the Hermitian conjugate of $\tilde{L}_{-1}$.
$\tilde{L}_{-1}=\sum_{n\geq 0} \tilde{a}_{-n-1}\tilde{a}_{n}=\sum_{n\geq 0} a_{-n-1}a_{n}\equiv L_{-1}$.
Here we have defined $L_{n}=\frac{1}{2}\sum_{m\in Z} :a_m a_{n-m}:$. It can be checked that $L_{-n}^\dagger=L_n$ and $L_0|p_{1,-1}\rangle =\frac{k^2}{2}|p_{1,-1}\rangle $
Thus the normalization of $J_{1^n}$ reads
\begin{eqnarray}
\langle p_{1,-1}|(L_1)^n
(L_{-1})^n| p_{1,-1}\rangle&=&(2h+n-1)(n)(2h+n-2)(n-1)\cdots(2h)\cdot
1\\\nonumber&=&\prod_{s\in
1^n}(l(s)+1+a(s)\frac{1}{\beta})(l(s)+(a(s)+1)\frac{1}{\beta}) \end{eqnarray}
which coincides with the Stanley's normalization for the Jack
polynomials \cite{Stanley}. Since there is a natural duality in CS model which states that
if one change $k\rightarrow-1/k$ and meanwhile transpose the
partition(Young tableau), the theory doesn't change. This implies
one can define the Jack polynomial with Young tableau $\{n\}$ as:
\[\langle k|(L_1)^n\] up to a normalization factor $k^{-2n}$,
\[J_{n}=k^{-2n}\langle k|(L_1)^n=n!k^{-2n}\langle 0|\oint
e^{-k\varphi(w)}w^{n-1} dw .\]

\section{Skew-Recursion Formula for Jack States}
In the previous section we have shown that any ``Virasoro'' null vector, represented by a  multiple integral of the Selberg type, is a Jack state of the rectangular graph up to normalization. One may naturally ask how the other Jack states be represented. Our answer to this question is positive.  In this and the following sections we shall show that any ``Virasoro'' null vector, or equivalently, the Jack state of the rectangular graph, skewed by  another Jack state is again a Jack state. In this way we can build any desired Jack state recursively either in operator or multiple integral formalism.
There are already two kinds of integral representations of the Jack
symmetric polynomials\cite{Awata:1995ky}\cite{Okounkov97a}.
Both are based on the method that the
number of arguments $N$ in $J_\lambda(p)$ are increased recursively.
The method we have developed is, however, in a different manner.
While other methods are based on adding blocks of squares to the Young tableau,
we are trying to subtract a block of squares from a given rectangular one. And
the other difference is that we first build an operator formalism,
and later an integral formalism based on it (in contrast to the pure operator formalism,
\cite{Lapointe(1995)}.
The way to subtract a block of squares from a given Young tableau is described in mathematical language as "skewing". We have already seen this method in section 2.5. The skewing of $\lambda$ by $\mu$ when $\lambda$
is a rectangular one is, however, simpler. In this case, the
summation only contains one term. This fact is proven by Kadell in
\cite{Kadell97} and is presented as $P_{\lambda}(p)\prod_{i=1}^{N}z_i^{-n}=P_{n^N/\lambda}(p^\ast)$ with the Young tableau
$n^N/\lambda:=\{n,\cdots, n, n-\lambda_{l}, n-\lambda_{l-1},\cdots ,n-\lambda_{1}\}\Rightarrow \lambda_1\leq n$ and $l\leq N$. In fact, in eq.(\ref{fusioncoeff}), we have made use of this identity  in the calculation of the fusion coefficients. Here, however, we shall
show that this particular skew relation has profound meaning related
to the Virasoro singular vectors. One can also view our method as an alternative proof
on Kadell's formula, eq.(\ref{Kadell}).

\subsection{Proposition and Examples}

\figscaled{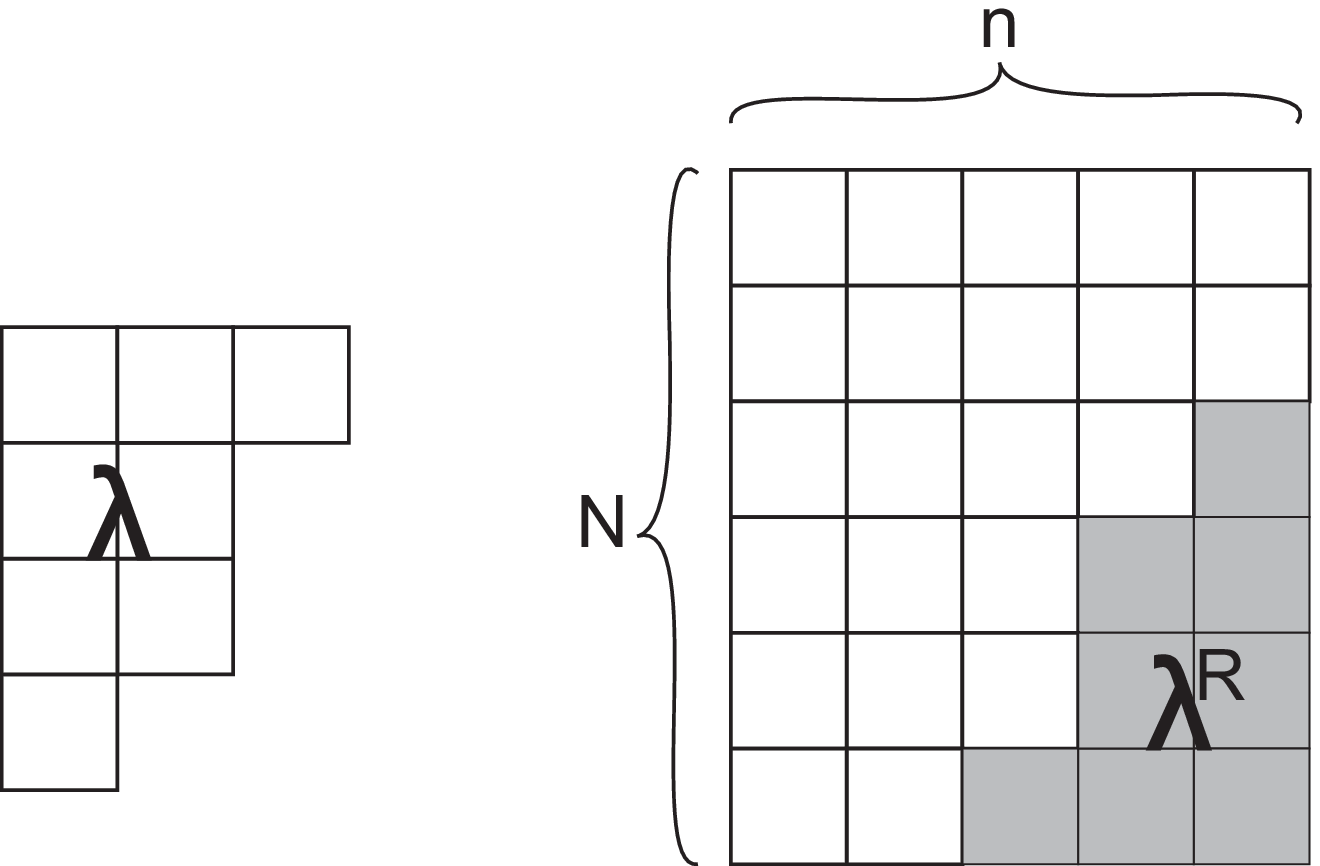}{The Young tableau for $n^N/\lambda$, the shadowed part labeled as $\lambda^R$ has been cut out from $n^N$.}{height = 2in}
\begin{proposition}
 Given a  Jack bra state of the rectangular graph,
\[|p_{-N,-n}\rangle_{\{n^N\}}=J_{-n^N}|p_{-N,-n} \rangle,\] if it is acted from the left by a
Jack annihilation operator  $J_{\lambda}$, $\lambda \prec n^N$,
$J_{n^N/\lambda}|p_{-N,-n}\rangle:=J_{\lambda}( \frac{a}{k} )|p_{-N,-n} \rangle_{\{n^N\}}$, then $J_{n^N/\lambda}|p_{-N,-n}\rangle$
is again a Jack bra state up to a normalization constant.
\end{proposition}

\[J_{\lambda}(\frac{a}{k} )|p_{-N,-n} \rangle_{\{n^N\}}\propto|p_{-N,-n} \rangle_{\{n^N/\lambda\}}.\] Here, the introducing of $p_{-N,-n} $ for the oscillator vacuum state is artificial. It just make the comparison with the ``Virasoro'' null vector easier. The Young
tableau $\{n^N\}/\lambda$ is shown in fig.\ref{skewyoung}. Before
rushing to the proof of the proposition, we start from some simple examples according to the level of the graphs being cut.
\subsubsection{Example 0: level 0}
In order to show that
\be\label{level0}
|p_{-N,-n}\rangle_{\{n^N\}} \propto |\chi_{N,n}\rangle\,,
\ee
we just have to calculate
\bea
\hat{H} |\chi_{N,n}\rangle &=&(N\beta+\beta-1-\alpha_+\tilde{a}_0)\tilde{a}_{-n}\tilde{a}_n|\chi_{N,n}\rangle \\\nonumber
 &=&N n^2 |\chi_{N,n}\rangle
\eea
Notice that $E_{n^N}=N n^2$, eq.(\ref{level0}) is proved.

\subsubsection{Example 1: level 1}
Level one graph is just a single square. If we cut a square in the SE corner of
the rectangular graph, the resulting state is proportional to \[\tilde{a}_1J_{n^N}|p_{-N,-n} \rangle.\]
It is easy
to show that the "Virasoro part" of the Hamiltonian have
eigenvalue on the resulting state
\[\alpha_+\tilde{a}_{-n}\tilde{L}_n\tilde{a}_1|p_{-N,-n} \rangle_{\{n^N\}}=
((N-1)\beta-(n-1))\tilde{a}_1|p_{-N,-n} \rangle_{\{n^N\}},\]
And the diagonal part of $\hat{H}$ has the eigenvalue \[(\beta N+\beta-1-\alpha_+\tilde{a}_{0})(nN-1)\tilde{a}_1|p_{-N,-n} \rangle_{\{n^N\}}=n(nN-1)\tilde{a}_1|p_{-N,-n} \rangle_{\{n^N\}}.\] Combining the two parts together, we have \[ \hat{H}\tilde{a}_1|p_{-N,-n} \rangle_{\{n^N\}} =E_{n^N/\Box}\tilde{a}_1|p_{-N,-n} \rangle_{\{n^N\}},\]
Again this state  has the correct property
corresponding to the skew Young tableau $\{n^N/\Box\}$.
\subsubsection{Example 2: level 2}
There are two different Young tableaux $\lambda^{(1)}$ and $\lambda^{(2)}$ at level 2. If we cut these Young tableaux  from a rectangular one $s^r$, the resulting states will span a two dimensional Hilbert space. Let us denote them as
\[|\chi\rangle=(\frac{\tilde{a}_1^2}{2k^2}+A\tilde{a}_2)|\psi\rangle_{\{n^N\}},\]
here $A$ is an undetermined parameter. Note that the ``diagonal part'' of
the Hamiltonian only shift the eigenvalue by a global constant. So for the eigen-equation
\[\hat{H}|\chi\rangle=E_{\chi}|\chi\rangle.\] we can drop this diagonal
term and  consider only the "Virasoro part" of the Hamiltonian. After
a simple computation, one finds
\[A=\frac{1}{\sqrt{2}k^3} \,\,\,\text{or}\,\,\, \frac{-1}{\sqrt{2}k}\] corresponding to
$\lambda^{(1)}=\{2\}$ and $\lambda^{(2)}=\{1^2\}$ respectively.
\subsubsection{Example 3: level 3}
It is straightforward to continue on to level 3 graphs being cut. The resulting state is denoted as
\[|\chi\rangle=(\frac{\tilde{a}_1^3}{2\sqrt{2}k^3})+A\tilde{a}_1\tilde{a}_2+B\tilde{a}_3|\psi\rangle_{\{n^N\}},\]here
$A ,B$ are undetermined parameters. There are three independent
solutions for the eigen-equation corresponds to the three Young
tableau at level 3.

For the horizontal Young tableau $\{3,0\}$, one gets :
\(A=3/2k^4,\,\,\,\,B=\sqrt{2}/k^5.\)

For the vertical Young tableau
$\{1,1,1\}$, \(A=-3/2k^4,\,\,\,\,B=\sqrt{2}/k^2.\)

For the symmetric Young tableau
$\{2,1\}$, \(A=-\frac{1}{2k^2}(1-\frac{1}{k^2}),\,\,\,\,B=\frac{-1}{\sqrt2
k^3}.\)

These results reproduce the level 3 Jack polynomials.

\subsection{Proof by Brute Force Operator Formalism}
Having checked for the low level skew Jack states, we are encouraged to find a more general proof for the proposition 1.
In this section, we shall show that if $\langle J_\lambda|$ is a Jack
symmetric function related to Young tableau $\lambda$, then
$J_\lambda|\chi_{r,s}\rangle$ is proportional to a Jack symmetric function
related to a Young tableau $s^r/\lambda$, with $\lambda \prec s^r$. Here,
$|\chi_{r,s}\rangle$ is a Virasoro singular vector descendant from $|p_{-r,-s}\rangle$,
see eq.(\ref{recjackint}). We can prove this in operator formalism first by ``brute
force''. Later in the next section we shall present it in a more
compact manner. To proceed, we need to write the operator valued Jack function as follows: \be J_{\lambda} =
\sum_{\lambda',|\lambda'|=|\lambda|}C_{\lambda}^{\lambda'}a_{\lambda'_1}\cdots
a_{\lambda_s'}\,.\ee Then consider the commutator of $J_{\lambda}$ and $\hat{H}$ defined in eq.(\ref{OperatorH}),
\beaa\label{JH1}[J_\lambda, \hat{H}] &=&
\sum_{\lambda',l}C_{\lambda}^{\lambda'}a_{\lambda_1'}\cdots[a_{\lambda'_l},
\hat{H}]\cdots a_{\lambda'_s}
\\&=&\sum_{ \lambda',l}C_{\lambda}^{\lambda'}a_{ \lambda_1'}\cdots [(1-\beta) (\lambda'_l)^2 a_{\lambda'_l}+
2k \lambda'_l L_{\lambda'_l}'+\beta Nl a_{\lambda'_l}]\cdots
a_{\lambda_s'}\,,\eeaa here $$L'_l = \frac{1}{2}\sum_{m\in Z}(:a_m a_{l-m}:) - a_0
a_l\,.$$ In deriving this, we have used the commutation between $a_l$ and
$\hat{H}$: \bea[a_l,\hat{H}] &=& (1-\beta)l^2 a_l+\sum_{m>0} 2k l
a_{-m}a_{l+m} +\sum_{l>m>0}k l a_{l-m}a_m +\beta N l a_l
\\\nonumber&=& (1-\beta)l^2 a_l +2kl L'_l +\beta N l a_l\,.\eea
In moving $L_{\lambda'_l}'$ to the most left by the commutation relation $[L'_n, a_m]=-m a_{m+n}$ for $n,m>0$, more terms are generated,
\bea\label{3terms}[J_{\lambda}, \hat{H}] &=&
\sum_{\lambda',l}C_{\lambda}^{\lambda'}a_{\lambda_1'}\cdots
a_{\lambda'_{l-1}}[(1-\beta)(\lambda'_l)^2 a_{\lambda'_l} +\beta N \lambda'_l
a_{\lambda'_l}]\cdots a_{\lambda'_s} \\\nonumber&+&
\sum_{\lambda',n<l} C_{\lambda}^{\lambda'}(2k \lambda'_l
\lambda'_n) a_{\lambda_1'} \cdots a_{\lambda'_n+\lambda'_l} \cdots a_{\lambda'_{l-1}}a_{\lambda'_{l+1}}\cdots
a_{\lambda_s'}\\\nonumber&+&\sum_{\lambda',l}C_{\lambda}^{\lambda'}(2k
\lambda'_l)L_{\lambda_l'}'a_{\lambda_1'}\cdots a_{\lambda_{l-1}'}a_{\lambda'_{l+1}}\cdots
a_{\lambda'_s}\,.\eea Let us define some notations to simplify
our calculation. We denote the first line on the r.h.s. of eq.(\ref{3terms}) as $A_0$ since this term retain the same number of $a_n$'s comparing to the original terms in $J_{\lambda}$, the second line is named  as
$A_{-}$ since it contains one less $a_n$ comparing to the original term in
$J_{\lambda}$, the third line  separates into two terms
$A_++\slashed{A}$, which are defined as \beaa A_+ &=&
\sum_{\lambda',l,\lambda'_l>m>0} C_{\lambda}^{\lambda'} k \lambda'_l a_{\lambda_1'}\cdots
a_{\lambda_{l-1}'}
(a_{\lambda'_{l}-m}a_m )a_{\lambda'_{l+1}}\cdots a_{\lambda'_s}\\
\slashed{A} &=& \sum_{\lambda',l,m>0} C_{\lambda}^{\lambda'} (2k \lambda'_l)
(a_{-m} a_{\lambda'_{l}+m})a_{\lambda_1'}\cdots
a_{\lambda_{l-1}'}a_{\lambda'_{l+1}}\cdots a_{\lambda_s'}\,.\eeaa
If we apply eq.(\ref{3terms}) to a bra vacuum sate $\bra{0}$, the
contribution of $\slashed A$ vanishes. Since $\bra{J_\lambda}$ is an eigenstate of $\hat{H}$, we conclude
\[
\bra{0}J_\lambda \hat{H} = \bra{0} J_\lambda E_\lambda\Rightarrow
\]
\be\label{SA}
\left[J_\lambda,\hat{H}\right] = E_\lambda J_\lambda +\slashed{A}
\ee
\be\label{3A}
E_{\lambda}J_{\lambda} = A_+ +A_- +A_0
\ee
Now we calculate the action of $\hat{H}$ on the ket state $J_{\lambda}\ket{\chi_{r,s}}$
\be\label{eigenket}
\hat{H} J_{\lambda}\ket{ \chi_{r,s}} =
[\hat{H},J_{\lambda}]\ket{ \chi_{r,s}} + J_{\lambda}\hat{H}
\ket{ \chi_{r,s}}=[\hat{H},J_{\lambda}]\ket{ \chi_{r,s}} + E_{rs}J_{\lambda}
\ket{ \chi_{r,s}}\,.
\ee
By moving $L_{\lambda_l}'$ in eq.(\ref{JH1}) to the most right, we get
\bea\label{rightact}[\hat{H}, J_{\lambda}] &=& E_{\lambda} J_{\lambda}-
2A_+ -2A_0
\\\nonumber &-& \sum_{\lambda',l}(2k\lambda'_l)C_{\lambda}^{\lambda'}
a_{\lambda_1'}\cdots a_{\lambda'_{l-1}}a_{\lambda'_{l+1}}\cdots
a_{\lambda'_{s}}\left[L_{\lambda'_l}'-\frac{1}{2}\sum_{\lambda'_l>m>0}a_{\lambda'_{l}-m}a_m\right]\,.\eea
and
\bea\label{rightact2}
&&2A_+ +2A_0 +
\sum_{\lambda',l}(2k\lambda'_l)C_{\lambda}^{\lambda'} a_{\lambda_1'}\cdots
a_{\lambda'_{l-1}}a_{\lambda'_{l+1}}\cdots
a_{\lambda'_{s}}\left[L_{\lambda'_l}'-\frac{1}{2}\sum_{\lambda'_l>m>0}a_{\lambda'_{l}-m}a_m\right]\\\nonumber
&=& \sum_{\lambda',l}2k \lambda'_l C_{\lambda}^{\lambda'}
a_{\lambda_1'}\cdots a_{\lambda'_{l-1}}a_{\lambda'_{l+1}}\cdots
a_{\lambda'_s}\\\nonumber&\times& [ \sum_{m>0}(a_{-m} a_{\lambda'_l +m}) +\sum_{\lambda'_l>m>0}
(a_{\lambda'_l-m}a_m)
 +[-(k-\frac{1}{k})\lambda_l' +N
k]a_{\lambda'_l}] \\\nonumber
&=& 2k(\sqrt{2}\alpha_0 - \sqrt{2}\tilde{a}_0)+ Nk)|\lambda|J_\lambda +\sum_{\lambda',l}2k \lambda'_l C_{\lambda}^{\lambda'}
a_{\lambda_1'}\cdots a_{\lambda'_{l-1}}a_{\lambda'_{l+1}}\cdots a_{\lambda'_s} \tilde{L}_{\lambda'_l}
\,,\eea
In deriving these, use has been made of eq.(\ref{3A}) and eq.(\ref{Virredef}).
Substituting the results in eqs.(\ref{rightact}-\ref{rightact2}) to eq.(\ref{eigenket})
and using the property of the Virasoro singular vector,  $\tilde{L}_l\ket{\chi_{r,s}} = 0, \, \, l>0$, we conclude
\bea\label{bruteqn}
\hat{H}J_{\lambda}\ket{\chi_{r,s}} = \left[ E_{\lambda}+E_{r,s}
-|\lambda|\hat{M}\right]J_{\lambda}\ket{\chi_{r,s}}\,,\eea here
$$\hat{M} = 2k ( \sqrt{2}\alpha_0 - \sqrt{2}\tilde{a_0}+ N k),$$  on $\ket{\chi_{r,s}}$, $\tilde{a_0}$
gives \be\label{a0}
\tilde{a}_0\ket{\chi_{r,s}} = \left(\dfrac{1+r}{2}\alpha_+ +
\dfrac{1+s}{2}\alpha_-\right)\ket{\chi_{r,s}}\, .\ee
The establishment of eq.(\ref{bruteqn}) finishes the proof of the proposition 1 we proposed before, that is, Jack polynomials for rectangular Young
tableaux, skewed by an  Jack state is again a Jack state.

\subsection{More Compact Proof}
The proposition 1 is proven in the previous subsection by making use of the
eigen-equation for $\hat{H}$. However, we know that the eigenstate  of  $\hat{H}$
can always been written as an integral transformation,
\bea\label{intKernel}\bra{J_\lambda} &\propto& \bra{0} \oint
e^{k\sum_{n>0}\frac{a_{n}}{-n}p_{-n}} \prod_{i<j}|z_i - z_j|^{2\beta}
J_{\lambda}(\{z_i\})\prod_i \dfrac{dz_i}{z_i}\\\nonumber
&\equiv&\bra{0}\oint F_{\lambda}(a,z)\prod_i \dfrac{dz_i}{z_i}\,. \eea
Here $a\equiv \{a_n\}$, $z\equiv \{z_i\}$, and in the following the integration measure $\prod_i \dfrac{dz_i}{z_i}$ will be implied without written out explicitly.
With $J_\lambda$ realized in this way, we found that the brute force proof can be rewritten in a more compact form with less indices involved.
Using: \bea
\label{usfulformula} a_{-m} e^{k \sum_{n>0}\frac{a_n}{-n}p_{-n}} =
e^{k\sum_{n>0}\frac{a_{n}}{-n}p_{-n}} (a_{-m} + kp_{-m})\\
e^{k\sum_{n>0}\frac{a_n}{-n}p_{-n}}a_{-m} =
(a_{-m}-kp_{-m})e^{k\sum_{n>0}\frac{a_n}{-n}p_{-n}}\,,\eea we have \bea\label{compact1}  \int
F_{\lambda}(a,z)\hat{H} &=& \int \left[\sum_{n,m=1}^{\infty}k\left((a_{-n} -
k p_{-n})(a_{-m}- kp_{-m}) a_{n+m} +(a_{-n-m}-k
p_{-n-m})a_na_m\right)\right.\\\nonumber&+&\left.
\sum_{n=1}^\infty(a_{-n}-kp_{-n})a_n(\beta N +
n(1-\beta))\right]F_{\lambda}(a,z)\\\nonumber&=& \int \hat{H} F_{\lambda}(a,z)
+ \int\left[ \sum_{n,m=1}^{\infty}(k^3
p_{-n}p_{-m}a_{n+m} -2k^2p_{-m}a_{-n}a_{n+m}  -k^2 p_{-n-m} a_n a_m)
\right.\\\nonumber&-&\left.\sum_{n=1}^{\infty}k p_{-n}a_n(\beta
N+n(1-\beta))\right] F_{\lambda}(a,z)\,. \eea Since the terms containing
$a_{-n}$'s on the most left will annihilate the bra vacuum $\bra{0}$,
we conclude the following identity \bea\label{moveeq}\left[J_\lambda,
\hat{H}\right]&=& E_\lambda J_\lambda - \int 2k^2 \sum_{n,m =1}^{\infty}p_{-m}
a_{-n} a_{n+m} F_{\lambda}(a,z)\eea will  be true. Comparing eq.(\ref{compact1}) and eq.(\ref{moveeq}), we have
\bea\label{compactE}
&&\int\left[\sum_{n,m=1}^{\infty}(k^3 p_{-n} p_{-m} a_{n+m} -k^2
p_{-n-m} a_n a_m) \right.\\\nonumber&-&\left.\sum_{n=1}^{\infty}k
p_{-n}a_n(\beta N +n(1-\beta))\right] F_{\lambda}(a,z)= E_\lambda J_\lambda\,.\eea

Now we move $a_{-n}$'s in the last term in eq. (\ref{moveeq}) to the
most right to get: \be\label{compactright} \left[J_\lambda, \hat{H}\right] = E_\lambda J_\lambda - \int 2k^2
\sum_{n,m=1}^{\infty}F_{\lambda}(a,z) p_{-m}(a_{-n}+kp_{-n})a_{n+m}\,.
\ee Using eq.(\ref{compactE}), the last term in the above eqation can be rewritten as
\bea && -2E_\lambda J_\lambda -\int
F_{\lambda}(a,z)\left( \sum_{n,m=1}^\infty 2k^2 (p_{-m}a_{-n}a_{n+m}  +
 p_{-n-m}a_n a_m) \right.\\\nonumber&+&\left. \sum_{n=1}^\infty 2k p_{-n} a_n(\beta N
+n(1-\beta))\right)\,,\eea
Substituting this result into eq.(\ref{compactright}), we have \bea\label{comutationHJ}\left[J_\lambda,
\hat{H}\right] &=& -E_\lambda J_\lambda -2k^2 \int F_{\lambda}(a,z)
\left\{\sum_{n,m=1}^\infty
p_{-m}\left(\sum_{n=1}^{\infty}(a_{-n}a_{n+m})+\sum_{n=1}^{m-1}a_n
a_{m-n}\right)\right.\\\nonumber &+& \left.\sum_{m=1}^\infty
p_{-m}a_m(kN + m(k^{-1}-k))\right\}\\\nonumber &=& -E_\lambda J_\lambda -2k^2
\int F_{\lambda}(a,z) \left\{\sum_{m=1}^\infty p_{-m}(\tilde{L}_m +
a_m(kN -(k^{-1}-k))-2a_0)\right\}\,,\eea where $\tilde{L}_m$ is the same as what we defined in eq.(\ref{Virredef}). When we  apply eq.(\ref{comutationHJ}) to a
Virasoro singular vector $ \ket{\chi_{rs}} $, $\tilde{L}_n\ket{\chi_{rs}}=0$ implies:
\bea\label{skewenergy}\left[\hat{H}, J_\lambda\right] \ket{\chi_{rs}} = \left(E_\lambda J_\lambda + 2k^2
\int F_{\lambda}(a,z)\sum_{m>0} p_{-m} a_m(k N +(k-k^{-1})-2a_0)\right)\ket{\chi_{rs}} \,.\eea Now we can
check, using eqs.(\ref{intKernel}-\ref{usfulformula}),
\[-k\int F_{\lambda}(a,z)\sum_{m>0}p_{-m}a_m =[J_{\lambda},
\sum_{m>0}a_{-m}a_m] = |\lambda|J_\lambda\,,\] which leads to
\be\hat{H} J_\lambda\ket{\chi_{rs}} = \left(E_\lambda+ E_{\chi_{rs}}-2k|\lambda|(kN
+k-k^{-1}-2 a_0)\right)J_\lambda\ket{\chi_{rs}}\,.\ee Here and before we have assumed
that Virasoro $\tilde{L}_n$ singular state $\ket{\chi_{rs}}$ is an eigenstate for CS
Hamiltonian $\hat{H}$ with eigenvalue $E_{\chi_{rs}}$. This can be checked as follows. From the
formula, eq.(\ref{modifiedH}) \[H = k\sum_{n=1}^\infty a_{-n}\tilde{L}_n +
\sum_{n=1}^\infty(\beta N +\beta -1 -2k a_0)a_{-n}a_n\,,\] we
arrive at: $E_{\chi_{rs}} = (\beta N+\beta -1 -\sqrt{2}k p_{-r,-s})l$, here $l$ is the level of the decendant states. By the
construction of Virasoro singular vectors, we know $l=rs, \, \, \sqrt{2}p_{-r,-s} = (1+r)k-
(1+s)k^{-1}$, hence
 \bea
&&E_{\chi_{rs}} = \left[\beta N +\beta -1
-k((1+r)k-(1+s)\frac{1}{k})\right]|\lambda|\\\nonumber&& = rs^2
+\beta (N-r)rs = E_{\{s^r\}}\,.\eea Thus eq.(\ref{skewenergy})
implies that $J_\lambda\ket{\chi}$ is an eigenstate of $\hat{H}$ with the
eigenvalue \bea &&E_\lambda +E_{s^r} -2k |\lambda|(kN
+k-k^{-1}-(1+r)k+(1+s)k^{-1})\\\nonumber &=& E_\lambda + E_{s^r}
-2|\lambda|((N-r)\beta +s)) = E_{s^r/\lambda}\,.\eea This concludes
our proof of proposition 1.

\section{Skew-Recursive Construction of Jack States}
 In the previous sections we have shown that if we cut, inside a
rectangular Young tableau of size $r\times s$, any sub-Young tableau
in a skew way, the resulting Young tableau is unique and hence the
corresponding Jack function, which is named as $J_{s^r/\lambda}$. This
Jack function, $J_{s^r/\lambda}$, can be used again to cut another bigger
rectangular Young tableau of size $r_1\times s_1$ to get
$J_{s_1^{r_1}/(s^{r}/\lambda)}$ and so forth. If we know the construction
of the Jack function for a definite Young tableau, we can build a
tower of Jack functions upon it in such a skew way. Of course, if we start
with a trivial Young tableau (empty), then the tower of Jack
functions is built upon the constructions of the Jack function for
rectangular Young tableau only, which are in fact Virasoro singular vectors. Following is the precise procedure
which leads to the recursive construction of the Jack functions.

\subsection{Operator Formalism}

First, $J_{-\lambda}$ acts on the left
vacuum to create a bra state
\[_{\lambda}\bra{0} \equiv \bra{0}J_{\lambda} ,\] $J_{\lambda}$ acts to the right will
produce a skew ket state \bea \label{skewstate}J_{\lambda}
\ket{0}_{\{s_1^{r_1}\}} &\equiv& J_{\lambda}J_{-s_1^{r_1}}\ket{0}\equiv
J_{-s_1^{r_1 }/\lambda}\ket{0} \\\nonumber &=&
g_{\lambda,s_1^{r_1}/\lambda}^{s_1^{r_1}}
j^{-1}_{[\lambda,s_1^{r_1}]}J_{-[\lambda, s_1^{r_1}]}\ket{0}\\\nonumber
&=&g_{\lambda,s_1^{r_1}/\lambda}^{s_1^{r_1}}j^{-1}_{[\lambda, s_1^{r_1}]}
\ket{0}_{\{s_1^{r_1}/\lambda\}} \,.\eea
 Here we use the symbol $[\lambda,s^r]$
to represent the unique Young tableau $s^r/\lambda$, see fig.\ref{skewyoung}, where $\lambda^R$
means $\lambda$ rotated by $\pi$ angle. Such type of Young tableau, i.e.,
a rectangular one cut in the SE corner by a rotated $\lambda$, will be frequently used recursively. For example,
$[[\lambda,s_1^{r_1}],s_2^{r_2}]$ will define another Jack function
associated with the Young tableau $s_2^{r_2}$ cut in the SE corner
by $[\lambda,s_1^{r_1}]$ rotated.

To facilitate such recursive procedure, we shall define the
following abbreviation \bea [r,s]_{\lambda,n} &\equiv&
\left[\cdots[[ \lambda,
s_1^{r_1}],s_2^{r_2}],\cdots,s_n^{r_n}\right]\\\nonumber \langle
J_{(r,s)_{\lambda,n+1}}&\equiv& \langle
J_{s_{n+1}^{r_{n+1}}}J_{-(r,s)_{\lambda,n}}\,,\\\nonumber
J_{(r,s)_{\lambda,0}}&=& J_{\lambda}\,.\eea Here and after, however, we
shall take $\lambda$ to be the empty Young tableau, so we shall use the
abbreviation \bea [r,s]_n &\equiv&
\left[\cdots[[s_1^{r_1},s_2^{r_2}],s_3^{r_3}],\cdots,
s_n^{r_n}\right]\\\nonumber \bra{J_{(r,s)_{n+1}}}&\equiv&
\bra{J_{s_{n+1}^{r_{n+1}}}J_{-(r,s)_n}}\\\nonumber  J_{(r,s)_0}&=&
1\, .\eea
It is clear that any regular Young tableau can be represented uniquely by
two integer vectors of dimension $n$ each, $[r,s]_n$, where $n-1$ is the total
number of skews for the  Young tableau considered according to our convention.
From the definition eq. ({\ref{skewstate}), we know that
$J_{(r,s)_n}$ differ from the standard Jack symmetric function
$J_{[r,s]_n}$ only by a normalization constant. For
example, \bea |J_{-(r,s)_1}\rangle &=&
 J_{-s_1^{r_1}}\ket{0}\\ \bra{J_{(r,s)_2}}&=&
 \bra{0}J_{s_2^{r_2}}J_{-s_1^{r_1}}\equiv \bra{0}J_{s_2^{r_2}/s_1^{r_1}}
 \\\nonumber &=& g_{s_1^{r_1},s_2^{r_2}/s_1^{r_1}}^{s_2^{r_2}}j^{-1}_{s_2^{r_2}/s_1^{r_1}}\bra{0}J_{[s_1^{r_1},s_2^{r_2}]}
 \\ J_{-(r,s)_3}\ket{0} &=& J_{(r,s)_2}J_{-s_3^{r_3}}\ket{0}\\\nonumber&=&
 g_{s_1^{r_1},s_2^{r_2}/s_1^{r_1}}^{s_2^{r_2}}j^{-1}_{[s_1^{r_1},s_2^{r_2}]}J_{[s_1^{r_1},s_2^{r_2}]}J_{-s_3^{r_3}}\ket{0}
 \\\nonumber&=&  g_{s_1^{r_1},s_2^{r_2}/s_1^{r_1}}^{s_2^{r_2}}j^{-1}_{[s_1^{r_1},s_2^{r_2}]}
 g_{[s_1^{r_1},s_2^{r_2}],
 [[s_1^{r_1},s_2^{r_2}],s_3^{r_3}]}^{s_3^{r_3}}\\\nonumber &\times&
 j^{-1}_{[[s_1^{r_1},s_2^{r_2}],s_3^{r_3}]}J_{-[[s_1^{r_1},s_2^{r_2}],s_3^{r_3}]}\ket{0}
  \eea

In general, the normalization constant can be determined as following.
Suppose $$J_{(r,s)_n} =
C_{[r,s]_n}J_{[r,s]_n}\,,$$ then \bea
\bra{J_{(r,s)_{n+1}}} &=& \bra{0} J_{s_{n+1}^{r_{n+1}}}J_{-(r,s)_n}
= \bra{0} J_{s_{n+1}^{r_{n+1}}}J_{-[r,s]_n}
C_{[r,s]_n}\\\nonumber&=& g_{[r,s]_n,
[[r,s]_n,s_{n+1}^{r_{n+1}}]}^{s_{n+1}^{r_{n+1}}}j^{-1}_{[[r,s]_n,s_{n+1}^{r_{n+1}}]}
\bra{J_{[r,s]_{n+1}}}C_{[r,s]_{n+1}}\,,
\eea so $C_\lambda$ can be defined recursively: \be
C_{[r,s]_{n+1}} =
C_{[r,s]_n}g_{[r,s]_n,
[r,s]_{n+1}}^{s_{n+1}^{r_{n+1}}}
j_{[r,s]_{n+1}}^{-1},\ee
where the fusion coefficient $g_{[r,s]_n,
[r,s]_{n+1}}^{s_{n+1}^{r_{n+1}}}$ is calculated in eq.(\ref{coeff1}).

\subsection{Integral Representation}
In practice, an integral formalism is more useful in analysis. Based
on the operator formalism, we derive the following integrals for
building the Jack symmetric functions.

\subsubsection{Auxiliary Scalar Fields}
 Since $J_{s^r}$'s are
essentially the building blocks for any generic Jack function
$J_{[r,s]_n}$, we come back to the construction of
$J_{s^r}$ by the following integral,
\bea J_{-s^r}\ket{p}&=&
\int [dz]_r^+ \prod_{i=1}^r
z_i^{-s-1}e^{\sum_{n>0}\frac{a_{-n}p_n}{n}}\ket{p}\,,\eea here
we have defined
\beaa [dz]_r^+ &\equiv& \dfrac{B_{s^r}}{\Gamma_r^2}\prod_{i<j}
|z_i-z_j|^{2\beta}\prod_{i=1}^r dz_i
\\\, [dz]_s^- &\equiv&
\dfrac{(-1)^{sr}A_{s^r}}{\Gamma_r^2} \prod_{i<j}|z_i-z_j|^{2/\beta}\prod_{i=1}^s dz_i
\,. \eeaa

To relate $J_{-s^r}\ket{p}$ to a Virasoro singular vector, we
introduce two scalar field, $\varphi^{(0)}$ and $\varphi^{(1)}$ to provide the right integration
measure $[dz]$,
\[\langle\varphi^{(i)}(z)\varphi^{(j)}(z')\rangle =
\delta_{ij}\log(z-z')\,,\] and define the vertex operator integral
$$V_{01}^{\pm} = \oint:e^{k^{\pm 1}(\varphi^{(0)} +\varphi^{(1)})(z)}:\,\,\,\,.$$ Clearly, $V_{01}^{\pm}$ is the screening charge for the
Virasoro algebra $L_n^{\pm}$ respectively. Here \beaa
T^{01,\pm}(z)&\equiv& \sum_{n\in\mathbb{Z}} L_n^{\pm} z^{-n-2}
\\ &=& \frac{1}{4}(\partial_z(\varphi^{(0)}+ \varphi^{(1)}))^2 \pm \frac{1}{2}(k-\frac{1}{k})\partial^2_z(\varphi^{(0)}+\varphi^{(1)})\,.\eeaa

Define \beaa \ket{\chi_{rs}^+} &=& (V_{01}^+)^r\ket{p_{r,-s}} \\
\bra{\chi^-_{rs}}&=& \bra{p_{-r,s}}(V_{01}^-)^s\,,\eeaa clearly we have
\beaa L_n^+\ket{\chi_{rs}^+} = 0,
\,\,\,n>0\\\bra{\chi_{rs}^-}L_{-n}^- = 0,\,\,\,n>0\,.\eeaa
However, to get $J_{s^r}$, we have to project out one of the two scalar
fields, say, $\varphi^{(0)}$ and from eq.(\ref{skew}) we get,
\bea J_{s^r}(\dfrac{a^{(1)}_-}{k})\ket{p}_1 \propto _0\bra{p}\chi_{rs}^+\rangle\\
_1\bra{p}J_{s^r}(\dfrac{a^{(1)}}{k})\propto \bra{\chi_{rs}^-}p\rangle_0\,,\eea so that
now $J_{\pm s^r}$ contains only  $a^{(1)}_{\pm n}$'s.

Now the Jack states read \bea |J_{-s^r}\rangle &=& \int
\prod_{i=1}^r z_i^{-s-1} [dz]_r^+ e^{k\sum_{n>0}\frac{a_{-n}^{(1)}p_n}{n}}\ket{p}_1 \equiv J_{-(r,s)_1} \rangle \\
\langle J_{s^r}| &=& _1\bra{p}\int
e^{\frac{1}{k}\sum_{n>0}\frac{a_{n}^{(1)}p_{-n}}{-n}}\prod_{i=1}^s z_i^{r-1}
[dz]_s^- \equiv \langle J_{(r,s)_1}\,, \eea here $\ket{p}_i$ is the vacuum
state (no oscillator excitations) for the
$\varphi^{(i)}$ scalar
\be
a_n^{(i)}\ket{p}_i = \delta_{n,0} p^{(i)}\ket{p}_i\,\, \,\,n\geq0
\ee
Notice that since
$a_{-n}^{(0)}$ has been projected out, $J_{-s^r}$ is no longer a null
vector for $L_n^+$. However, $J_{-s^r}$ is still a null vector for
the modified Virasoro generator $\tilde{L}_n$ constructed with $\varphi^{(1)}$
only, see, eq. (\ref{Virredef}).

\subsubsection{Bra and Ket States}
Now we shall specify how the bra state $\bra{p^+_{p_{r,s}}}$ and the ket state
$\ket{p^-_{p_{r,s}}}$ are labeled.

Since we have $L_{n}^{\pm}$ acts on ket-state and bra-state
respectively, so we have different screening charges for $L_n^{\pm}
$ respectively. $$\alpha^{++} = \sqrt{2}k, \,\,\, \alpha^{+-} =
-\sqrt{2}k^{-1}$$ for $L_n^+$, and $$\alpha^{-+} = -\sqrt{2} k,
\alpha^{--} = \sqrt{2}k^{-1}$$ for $L_n^-$. If we combine
$\varphi^{(i)}+\varphi^{(i+1)}$ into a single scalar, $$\varphi =
\dfrac{1}{\sqrt{2}}(\varphi^{(i)}+\varphi^{(i+1)}),$$ and \beaa
a_0\ket{p_{r,s}} &=& p^+_{r,s}\ket{p_{r,s}} \\ \bra{p_{r,s}} a_0&=&
\bra{p_{r,s}}p^-_{rs}\,, \eeaa then we define \bea p^+_{rs} &=&
\frac{1}{2}(1-r)\alpha^{++} + \frac{1}{2}(1-s)\alpha^{+-}\\\nonumber
&=& \frac{1}{2}(1-r)\sqrt{2} k-\frac{1}{2}(1-s)\frac{\sqrt{2}}{k}\\
p_{rs}^- &=& \frac{1}{2}(1+r)\alpha^{-+}+\frac{1}{2}(1+s)\alpha^{--}
\\\nonumber&=& -\frac{1}{2}(1+r)\sqrt{2}k +\frac{1}{2}(1+s)\frac{\sqrt{2}}{k}\eea

Now consider $$a_0\ket{p^+_{r,s}}_{i,i+1} = p^+_{r,s}\ket{p_{r,s}}_{i,i+1}\,.$$ However,
when, say $\varphi^{(i)}$ is projected out, then
$$a_0^{(i+1)}\ket{p^+_{r,s}}_{i+1} =
\dfrac{1}{\sqrt{2}}p^+_{r,s}\ket{p^+_{r,s}}_{i+1}\,.$$ For $\bra{p^-_{r,s}}$,
the projection is similar. To see this notation will provide the
correct integration measure, one could check: \bea
&&\bra{ p^-_{-r,s}}(V^-)^s/\Gamma_s^2 \ket{p^+_{-r,-s}}_i\\\nonumber&=& \bra{ p^-_{-r,-s}}
\int\dfrac{\prod_{i<j} (z_i-z_j)^{\frac{2}{k^2}}\prod_{i=1}^s z_i
^{\frac{\sqrt{2}}{k}a_0}}{\Gamma_s^2}
e^{\frac{\sqrt{2}}{k}\sum_{n>0}\frac{a_n}{-n}p_{-n}}\prod dz_i \ket{p^+_{-r,-s}}_i\\\nonumber
&=&
_{i+1}\bra{p^-_{-r,-s}}\int\dfrac{\prod_{i<j}(z_i-z_j)^{\frac{2}{k^2}}}{\Gamma_s^2}\prod_{i=1}^sz_i^{\frac{\sqrt
2}{k}(\frac{1}{2}(1-r)(-\sqrt 2 k)+\frac{1}{2}(1-s)\frac{\sqrt
2}{k})}e^{\frac{1}{k}\sum_{n>0}\frac{a^{(i+1)}_n}{-n}p_{-n}}\prod_i dz_i
\\\nonumber &=& _{i+1}\bra{ p^-_{-r,-s}} \int \prod_{i<j}\left[\dfrac{(z_i-z_j)^2}{z_i z_j}\right]^{\frac{1}{k^2}}
e^{\frac{1}{k}\sum_{n>0}\frac{a^{(i+1)}_n}{-n}p_{-n}}\prod_{i=1}^s z_i^{r-1}d
z_i/ \Gamma_s^2
\\\nonumber &=& _{i+1}\bra{\chi_{rs}} \propto \bra{ p^-_{-r,-s}}\int e^{\frac{1}{k}\sum_{n>0}\frac{a^{(i+1)}_n}{-n}p_{-n}}\prod_{i=1}^s (z_i)^{r-1}[dz]^-_s\\\nonumber
&=& _{i+1}\bra{p^-_{-r,-s}}J_{s^r}(\frac{a^{(i+1)}}{k}) \,,\eea
produces the Jack states of rectangular graph.

\subsubsection{Integral Recursion} Now we have \bea
\ket{J_{-s_1^{r_1}}} &=& \ket{J_{-(r,s)_1}} =
\,\,_0\bra{p_0}{\chi_{r_1,s_1}}\rangle_{01} \\\nonumber&=&
\int e^{\sum_{n>0}\frac{a_{-n}^{(1)}p_n}{n}k}\prod_{i=1}^{r_1}(z_{1,i})^{-s-1}[d
z_1]_{r_1}^+\ket{ p^+_{-r_1,-s_1}}_1\\\nonumber p_0 &=&
\frac{1}{\sqrt{2}}p^+_{-r_1,-s_1} =
\frac{1}{2}(1+r_1)k-\frac{1}{2}(1+s_1)\frac{1}{k}\,.\eea
\figscaled{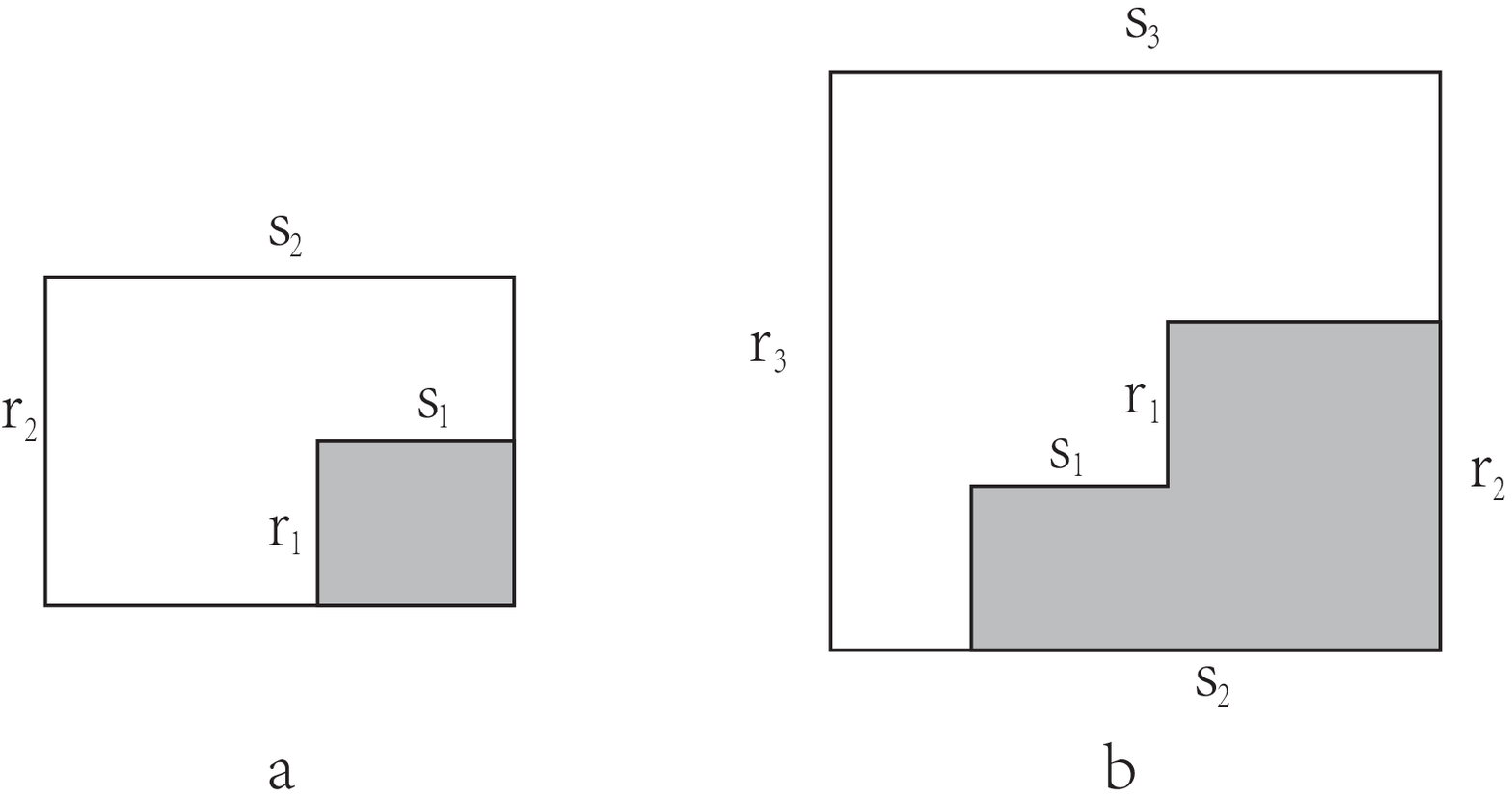}{a. Young tableau $\{s_2^{r_2}\} /\{s_1^{r_1}\}$
\ \ \ b. Young tableau  $\{s_3^{r_3}\}/(\{s_2^{r_2}\} /\{s_1^{r_1}\})$, this is a three-ladder Young tableau.}{height=2.5in} For one skew
Young tableau of the type as in fig.\ref{Young02}.a, we have to introduce $\varphi^{(2)}$ scalar and project out $\varphi^{(1)}$ scalar. The resulting state is actually the skew Jack state, as what has been shown in eq.(\ref{skew}); We proceed to construct \bea \bra{J_{(r,s)_2}} &=&
_{12}\bra{\chi_{r_2,s_2}}e^{\delta k_{21} q}J_{-(r,s)_1}|p^+_{-r_1,-s_1}\rangle_1\\\nonumber
&=& _{12}\bra{ p^-_{-r_2,-s_2}}(V_{12}^-)^{s_2} e^{\delta k_{21}
q^{(1)}}J_{-{(r,s)_1}} |p^+_{-r_1,-s_1}\rangle_1\\\nonumber &=& _2\bra{p^-_{-r_2,-s_2}}\iint
e^{\sum{n>0}\frac{1}{k}\frac{a_{n_2}^{(2)}\sum
z_{2,i}^{-n_2}}{-n_2}}\prod_{i=1}^{s_2}(z_{2,i})^{r_2-1}
\prod_{s_2,r_1}(1-\frac{z_1}{z_2})\prod_{i=1}^{r_1}(z_{1,i})^{-s_1-1}[d
z_2]_{s_2}^- [dz_1]^+_{r_1}\,. \eea Here we have defined
\[\prod^{s_m, r_n}(1-\frac{z_n}{z_m})\equiv \prod_{i = 1}^{s_m}\prod_{j=1}^{r_n} (1-\frac{z_{n,j}}{z_{m,i}})\,.\]and $e^{\delta k_{21}q}$ is
introduced to eliminate the charge deficit in $\varphi^{(1)}$
sector, that is \bea && _1\bra{ p^-_{-r_2,-s_2} } e^{\delta
k_{21}q^{(1)}}\ket{ p^+_{-r_1,-s_1}}_1\neq 0 \,. \eea will give the following equation, \bea
\dfrac{1}{\sqrt{2}}p^-_{-r_2,-s_2} &=& \delta k_{21} +\frac{1}{\sqrt
2} p^+_{-r_1,-s_1}\\
\delta k_{21} &=& (p^-_{-r_2,-s_2}-p^+_{-r_1,-s_1})\dfrac{1}{\sqrt
2}
\\\nonumber &=& \frac{1}{2}(\frac{1}{k}-k)-\frac{k}{2}(1+r_1-r_2) +\frac{1}{2k}(1+s_1-s_2)
\\\nonumber &=& \dfrac{1}{\sqrt 2}\left\{\alpha_0^- + p^+_{r_1-r_2,s_1-s_2}\right\}\\\nonumber
2\alpha_0^{\pm} &=& \alpha^{\pm+} + \alpha^{\pm-}\,.\eea

For two skew Young tableau,fig.\ref{Young02}.b, $\varphi^{(3)}$ is introduced and $\varphi^{(2)}$ eliminated.
 \bea\ket{J_{-(r,s)_3}} &=& _2\langle p^-_{-r_2,-s_2}|
J_{(r,s)_2}e^{\delta k_{23} q^{(2)}}\ket{\chi_{r_3,s_3}}_{23}
\\\nonumber &=& _2\langle p^-_{-r_2,-s_2}|J_{(r,s)_2}| e^{\delta
k_{23}q^{(2)}}(V_2^+)^{r_3}\ket{p^+_{-r_3,-s_3}}_{23} \\\nonumber &=& \int
\prod_{i=1}^{r_3}(z_{3,i})^{-s_3-1}[dz_3]_{r_3}^+ \prod^{s_2,
r_3}\left(1-\frac{z_3}{z_2}\right)\prod_{i=1}^{s_2}(z_{2,i})^{r_2-1}
[dz_2]_{s_2}^-\\\nonumber &\times&
\prod^{s_2,r_1}\left(1-\frac{z_1}{z_2}\right)\prod_{i=1}^{r_1}(z_{1,i})^{-s_1-1}
[dz_1]_{r_1}^+ \exp\left(k\sum_{n>0}\frac{a_{-n}^{(3)}}{n}\sum_i
z_{3,i}^n\right)\ket{p^+_{-r_3,-s_3}}_3\eea Similarly,  we have \bea
\delta k_{23} +\frac{1}{\sqrt 2}p^+_{-r_3, -s_3}& = &\frac{1}{\sqrt
2}p^-_{-r_2,-s_2} \\\nonumber \delta k_{23} &=& \frac{1}{\sqrt
2}(-\alpha_0^+ - p^+_{r_2-r_3,s_2-s_3})=\frac{1}{\sqrt 2}(\alpha_0^-
+ p^-_{r_3-r_2,s_3-s_2})\\\nonumber &=&
\frac{1}{2}(\frac{1}{k}-k)-\frac{k}{2}(1+r_3-r_2)+\frac{1}{2k}(1+s_3-s_2)\,.
\eea

In general, proceed recursively, we have, for $n$ odd
\bea\label{IntResult01} J_{-(r,s)_n} \ket{p^+_{-r_n,-s_n}}_n &=& _{n-1}\langle p^-_{-r_{n-1},-s_{n-1}}|
J_{(r,s)_{n-1}} e^{\delta
k_{n-1,n}q^{(n-1)}}\ket{\chi_{r_n,s_n}}_{n-1,n}\\\nonumber&=& \iiint
\exp\left({k\sum_{m>0}\dfrac{a_{-m}^{(n)}\sum_{i=1}^{r_n}z_{n,i}^m}{m}}\right)
\prod_{i=1}^{r_n}(z_{n,i})^{-s_n-1}\prod^{s_{n-1},r_n}(1-\frac{z_n}{z_{n-1}})\\\nonumber
&\times&\prod_{i=1}^{s_{n-1}}(z_{n-1,i})^{r_{n-1}-1}\prod^{s_{n-1},r_{n-2}}(1-\frac{z_{n-2}}{z_{n-1}})\cdots
\prod_{i=1}^{s_2}(z_{2,i})^{r_2-1}\\\nonumber
&\times&\prod^{s_2,r_1}(1-\frac{z_1}{z_2})\prod_{i=1}^{r_1}(z_{1,i})^{-s_1-1}[d
z]^o_{[n]!} \ket{p^+_{-r_n,-s_n}}_n \,.
\eea
Here \bea \delta k_{n-1,n} &=& \frac{1}{\sqrt
2}\left(-\alpha_0^+ - p^+_{r_{n-1}-r_n,
s_{n-1}-s_n}\right)\\\nonumber &=& \frac{1}{\sqrt
2}\left(\alpha_0^-+ p^-_{r_n-r_{n-1},s_{n}-s_{n-1}}\right)
\\\nonumber &=& \frac{1}{2}(\frac{1}{k}-k)-\frac{k}{2}(1+r_n-r_{n-1})+\frac{1}{2k}(1+s_n-s_{n-1})\,.\eea
For $n$ even, \bea\label{IntResult02} _n\bra{p^-_{-r_n,-s_n}} J_{(r,s)_n}&=&
\bra{\chi_{r_n,s_n}} e^{\delta
k_{n,n-1}q^{(n-1)}}J_{-{(r,s)_{n-1}}}|p^+_{-r_{n-1},-s_{n-1}}\rangle_{n-1}\\\nonumber&=& _n\bra{p^-_{-r_n,-s_n}} \iiint
\exp\left({\dfrac{1}{k}\sum_{m>0}\dfrac{a_{m}^{(n)}\sum_{i=1}^{s_n}z_{n,i}^{-m}}{-m}}\right)
\prod_{i=1}^{s_n}(z_{n,i})^{r_n-1}\prod^{s_n,r_{n-1}}(1-\frac{z_{n-1}}{z_n})\\\nonumber
&\times&\prod_{i=1}^{r_{n-1}}(z_{n-1,i})^{-s_{n-1}-1}\prod^{s_{n-2},r_{n-1}}(1-\frac{z_{n-2}}{z_{n-1}})\cdots
\prod_{i=1}^{s_2}(z_{2,i})^{r_2-1}\\\nonumber
&\times&\prod^{s_2,r_1}(1-\frac{z_1}{z_2})\prod_{i=1}^{r_1}(z_{1,i})^{-s_1-1}[d
z]^e_{[n]!}\,.\eea Here, \bea \delta k_{n,n-1} &=& \frac{1}{\sqrt
2}\left(-\alpha_0^+ - p^+_{r_n-r_{n-1},s_n-
s_{n-1}}\right)\\\nonumber &=& \frac{1}{\sqrt 2} (\alpha_0^- +
p^-_{r_{n-1}-r_n,s_{n-1}-s_n}) \\\nonumber &=&
\frac{1}{2}(\frac{1}{k}-k) -
\frac{k}{2}(1+r_{n-1}-r_n)+\frac{1}{2k}(1+s_{n-1}-s_n)\,. \eea The
integration measures are defined as following: for $n$ odd, \[[d
z]_{[n]!}^o \equiv [dz_1]_{r_1}^+ [dz_2]_{s_2}^-\cdots [d
z_n]_{r_n}^+\,.\] For $n$ even, \[[dz]_{[n]!}^e \equiv [d
z_1]_{r_1}^+[dz_2]_{s_2}^-\cdots [dz_n]_{s_n}^-\,.\]

Eq.(\ref{IntResult01}) and eq.(\ref{IntResult02}) are the main
results of our present work. \footnote{In fact, one can easily see that the distinguishment between even and odd skews is artificial.} It provides an integral representation
for any Jack symmetric function which, in our formalism, is labeled
by two integer vectors of dimension $n$ each, $(r,s)_n$.

The integral representation not only provide a useful tool in
analyzing problems involving Jack symmetric functions, but also give
an explicit construction of the Jack
symmetric functions in terms of free bosons.  It is also desirable to work out explicitly the Selberg
type multi-integrals appearing in eq.(\ref{IntResult01}) and
eq.(\ref{IntResult02}).

\subsection{Integral Representation for Jack Symmetric Polynomials} Having got the
integral representation for a general Jack symmetric function, it is
then straightforward to get the Jack symmetric polynomials in any
number $N$ of arguments $z_i$. Notice that in the following we shall
present the unnormalized Jack polynomials. However, the
normalization constants can be easily worked out.

First, let us consider $n$ even, thus \bea
J_{(r,s)_n}^{1/k^2}(\{z_i\})&\equiv&\langle
J_{(r,s)_n}\exp\left(k\sum_{m>0}\frac{a_{-m}^{(n)}}{m}\sum_{i=1}^N
z_i^m\right)\ket{p^+_n}_n\\\nonumber &=& \iiint \prod^{s_n,
N}(1-\frac{z}{z_n})\prod_{i=1}^{s_n}(z_{n,i})^{r_n-1} [d
z_n]_{s_n}^- \prod^{s_n,
r_{n-1}}(1-\frac{z_{n-1}}{z_n})\prod_{i=1}^{r_{n-1}}(z_{n-1,i})^{-s_{n-1}-1}[dz_{n-1}]_{r_{n-1}}^+\\\nonumber &\times& \prod^{s_{n-2},
r_{n-1}}(1-\frac{z_{n-1}}{z_{n-2}})\prod_{i=1}^{s_{n-2}}(z_{n-2,i})^{r_{n-2}-1}[dz_{n-2}]_{s_{n-2}}^-\cdots \\\nonumber &\times& \prod^{s_2,
r_1}(1-\frac{z_1}{z_2})\prod_{i=1}^{r_1}(z_{1,i})^{-s_1-1}[dz_i]_{r_1}^+\,.\eea And for $n$ odd, \bea
J^{1/k^2}_{(r,s)_n}(\{z_i^{-1}\}) &\equiv&_n \bra{p_{n}^-}
\exp\left(\frac{1}{k}\sum_{m>0}\frac{a_m^{(n)}}{-m}\sum_{i=1}^{N}
z_i^{-m}\right) J_{-(r,s)_n}\rangle \\\nonumber &=& \iiint
\prod^{N,r_n}(1-\frac{z_n}{z})\prod_{i=1}^{r_n}(z_{n,i})^{-s_n-1}[d
z_n]_{r_n}^+\prod^{s_{n-1},r_n}(1-\frac{z_n}{z_{n-1}})\prod_{i=1}^{s_{n-1}}(z_{n-1,i})^{r_{n-1}-1}[d
z_{n-1}]^-_{s_{n-1}} \\\nonumber &\times&
\prod^{s_{n-1},r_{n-2}}(1-\frac{z_{n-2}}{z_{n-
1}})\prod_{i=1}^{r_{n-2}}(z_{n-2,i})^{-s_{n-2}-1}
[dz_{n-2}]^+_{s_{n-2}}\cdots \\\nonumber &\times&  \prod^{s_2,
r_1}(1-\frac{z_1}{z_2})\prod_{i=1}^{r_1}(z_{1,i})^{-s_1-1}[d
z_{1}]_{r_1}^+\,.\eea Now $p_n^\pm$ can be easily worked out, \bea
p^+_n &=& \frac{1}{\sqrt 2}p^-_{-r_n,-s_n}= \frac{1}{\sqrt
2}\left(\frac{1}{2}(1-r_n)\alpha^{-+}+\frac{1}{2}(1-s_n)\alpha^{--}\right)\\\nonumber
&=& -\frac{k}{2}(1-r_n)+\frac{1}{2k}(1-s_n)\\p^-_n &=&
\frac{1}{\sqrt 2} p^{+}_{-r_n,-s_n} =
\frac{k}{2}(1+r_n)-\frac{1}{2k}(1+s_n)\,.\eea

\section{Acknowledgement}
This work is part of the CAS program "Frontier Topics in Mathematical Physics" (KJCX3-SYW-S03) and
is supported partially by a national grant  NSFC(11035008).


\begin{thebibliography}{99}


\bibitem{Awata:1995ky}
  H.~Awata, Y.~Matsuo, S.~Odake and J.~Shiraishi,
  ``Excited states of Calogero-Sutherland model and singular vectors of the
  W(N) algebra,''
  Nucl.\ Phys.\  B {\bf 449}, 347 (1995)
  [arXiv:hep-th/9503043].

\bibitem{Awata:1994fz}
  H.~Awata, Y.~Matsuo, S.~Odake and J.~Shiraishi,
  ``A Note on Calogero-Sutherland model, W(n) singular vectors and generalized
  matrix models,''
  Soryushiron Kenkyu {\bf 91}, A69 (1995)
  [arXiv:hep-th/9503028].

\bibitem{Awata:1994zr}
  H.~Awata, M.~Fukuma, Y.~Matsuo and S.~Odake,
  ``Representation theory of W(1+infinity) algebra,''

\bibitem{Awata:1994xd}
  H.~Awata, Y.~Matsuo, S.~Odake and J.~Shiraishi,
  ``Collective field theory, Calogero-Sutherland model and generalized matrix
  models,''
  Phys.\ Lett.\  B {\bf 347}, 49 (1995)
  [arXiv:hep-th/9411053].

\bibitem{Baker}
T.~H.~ Baker, P.~J.~ Forrester,
``The Calogero-Sutherland Model and Generalized
Classical Polynomials''
Commun.~ Math. ~Phys. 188,~ 175 ~ 216 (1997)

\bibitem{Iso:1994ui}
  S.~Iso and S.~J.~Rey,
  ``Collective field theory of the fractional quantum hall edge state and the
  Calogero-Sutherland model,''
  Phys.\ Lett.\  B {\bf 352}, 111 (1995)
  [arXiv:hep-th/9406192].

\bibitem{Azuma:1993ra}
  H.~Azuma and S.~Iso,
  ``Explicit relation of quantum hall effect and Calogero-Sutherland model,''
  Phys.\ Lett.\  B {\bf 331}, 107 (1994)
  [arXiv:hep-th/9312001].

\bibitem{Zamolodchikov82}
   Al.~ B. ~Zamolodchikov, ``Conformal symmetry in two-dimensional space: Recursion representation of conformal block, Teoret. Mat. Fiz., 73:1 (1987), 103-110


\bibitem{Zamolodchikov84}
  Al.~ B. ~Zamolodchikov, ``Conformal symmetry in two dimensions: An explicit recurrence formula for the conformal partial wave amplitude", Commun. Math. Phys. ~96, ~3, ~419-422,
\bibitem{Yamada95}
K.~ Mimachi and Y.~ Yamada ,``Singular vectors of the Virasoro
algebra in terms of Jack symmetric polynomials", Comm. Math. Phys.
~174, ~2 (1995), ~447-455 (1984)

\bibitem{Ha(1995)} Ha, Z.~N.~C.\ 1995, ``Fractional statistics in one dimension: view from an exactly solvable model'' Nuclear Physics
B, 435, 604

 \bibitem{Haldane07b} Bernevig, B.~A. and {Haldane}, F.~D.~M.,
``{Model Fractional Quantum Hall States and Jack Polynomials}",
Physical Review Letters, 2008, 100, 24,
arXiv: {0707.3637}

\bibitem{Okounkov97a}
A.~Okounkov and G.~Olshanki
``Shifted Jack Polynomials, Binomial Formula, and Applications"
Mathematical Research Letters 4, 69-78 (1997)

\bibitem{Lapointe(1995)} L.~Lapointe, \& L.~Vinet,\ 1995,~
 ``Exact operator solution of the Calogero-Sutherland model'', arXiv:q-alg/9509003

\bibitem{Polychronakos:1992zk}
  A.~P.~Polychronakos,
  ``Exchange operator formalism for integrable systems of particles,''
  Phys.\ Rev.\ Lett.\  {\bf 69}, 703 (1992)
  [arXiv:hep-th/9202057].

\bibitem{Sakamoto:2004rn}
  R.~Sakamoto, J.~Shiraishi, D.~Arnaudon, L.~Frappat and E.~Ragoucy,
  ``Correspondence between conformal field theory and Calogero-Sutherland
  model,''
  Nucl.\ Phys.\  B {\bf 704}, 490 (2005)
  [arXiv:hep-th/0407267].

\bibitem{Kadell92}
K.~W.~J.~Kadell, ``An integral for the product of two Selberg-Jack symmetric polynomials" ,Compositio Mathematica, 87 ~no. ~1 (1993), ~p. 5-43


\bibitem{Kadell97}
K.~W.~J.~Kadell, ``The Selberg-Jack Symmetric Functions ",~
Advances in Mathematics
130,~1997,~33-102


\bibitem{Stanley}
R.~P.~Stanley, ``Some Combinatorial Properties of Jack Symmetric Functions'', Advances in Mathematics ~77,~76-115,~(1989)

\bibitem{Dotsenko:1984ad}
  V.~S.~Dotsenko and V.~A.~Fateev,
  ``Four Point Correlation Functions and the Operator Algebra in the
  Two-Dimensional Conformal Invariant Theories with the Central Charge $c<1$,''
  Nucl.\ Phys.\  B {\bf 251}, 691 (1985).

\bibitem{Feigin:1981st}
  B.~L.~Feigin and D.~B.~Fuks,
  ``Invariant skew symmetric differential operators on the line and verma
  modules over the Virasoro algebra,''
  Funct.\ Anal.\ Appl.\  {\bf 16}, 114 (1982)
  [Funkt.\ Anal.\ Pril.\  {\bf 16}, 47 (1982)].

\bibitem{Felder:1988zp}
  G.~Felder,
  ``BRST Approach to Minimal Methods,''
  Nucl.\ Phys.\  {\bf B317}, 215 (1989).


\bibitem{AFLT}
  V.~A.~Alba, V.~A.~Fateev, A.~V.~Litvinov and G.~M.~Tarnopolsky,
  ``On combinatorial expansion of the conformal blocks arising from AGT
  conjecture,''
  arXiv:1012.1312 [hep-th].

\bibitem{AGT}
  L.~F.~Alday, D.~Gaiotto and Y.~Tachikawa,
  ``Liouville Correlation Functions from Four-dimensional Gauge Theories,''
  Lett.\ Math.\ Phys.\  {\bf 91}, 167 (2010)
  [arXiv:0906.3219 [hep-th]].

\bibitem{Dijkgraaf:2009pc}
  R.~Dijkgraaf and C.~Vafa,
  ``Toda Theories, Matrix Models, Topological Strings, and N=2 Gauge Systems,''
  arXiv:0909.2453 [hep-th].

\bibitem{Belavin:2011js}
  A.~Belavin and V.~Belavin,
  ``AGT conjecture and Integrable structure of Conformal field theory for
  c=1,''
  Nucl.\ Phys.\  B {\bf 850}, 199 (2011)
  [arXiv:1102.0343 [hep-th]].

\bibitem{Nekrasov:2009rc}
  N.~A.~Nekrasov and S.~L.~Shatashvili,
  ``Quantization of Integrable Systems and Four Dimensional Gauge Theories,''
  arXiv:0908.4052 [hep-th].

\bibitem{Carlsson-Okounkov}
E.~Carlsson and A.~Okounkov,
``Exts and Vertex Operators,''
  arXiv: 0801.2565v2.

\bibitem{Donagi:1995cf}
  R.~Donagi and E.~Witten,
  ``Supersymmetric Yang-Mills theory and integrable systems,''
  Nucl.\ Phys.\  B {\bf 460}, 299 (1996)
  [arXiv:hep-th/9510101].

\bibitem{ShouWuYu2011}
B.~Shou, J.~F. ~Wu and M.~Yu, ``AGT conjecture and AFLT states: A complete
construction", to appear.

\bibitem{WuXuYu2011c}
J.~F.~Wu, Y.~Y.~Xu and M.~Yu, in preparation.

\bibitem{Macdonald}
I.~G.~Macdonald, Symmetric Functions and Hall Polynomials,~1995,~2nd Edition,~Clarendon Press Oxford
\end{thebibliography}
 \end{document}